\documentclass[reprint,aps,prx,amsmath,amssymb,superscriptaddress,longbibliography,english
]{revtex4-2}
\pdfoutput=1
\usepackage{hyperref}

\usepackage{tikz}
\usepackage{tikz-cd}
\usepackage[shortlabels]{enumitem}
\usepackage{graphicx,color}
\usepackage{amsmath}
\usepackage{bm}
\usepackage{amssymb}
\usepackage[all]{hypcap}
\usepackage{mathtools}
\usepackage{dsfont}
\usepackage{physics}
\usepackage{braket}
\usepackage{enumitem}
\usepackage{float}
\usepackage{xcolor}
\usepackage{siunitx}

\renewcommand{\paragraph}[1]{\vspace{3pt}\textit{\textbf{#1}}.}
\makeatletter
\let\newfloat\newfloat@ltx
\makeatother

\begin{document}
\title{Dynamical Decoupling using Universal Optimal Tracking}

\author{Amit Devra}
\email{amit.devra@tum.de} 
\altaffiliation[Present address: ]{Munich Quantum Valley gGmbH, Walther-von-Dyck-Stra{\ss}e 6, 85748 Garching, Germany}
\affiliation{Technical University of Munich, TUM School of Natural Sciences, Garching 85748, Germany}
\affiliation{Munich Center for Quantum Science and Technology (MCQST), Schellingstra{\ss}e 4, M\"{u}nchen 80799, Germany}

\author{Emanuel Malvetti}
\altaffiliation[Present address: ]{Zurich Instruments AG, Technoparkstrasse 1, 8005 Zurich, Switzerland}
\affiliation{Technical University of Munich, TUM School of Natural Sciences, Garching 85748, Germany}
\affiliation{Munich Center for Quantum Science and Technology (MCQST), Schellingstra{\ss}e 4, M\"{u}nchen 80799, Germany}

\author{Niklas J. Glaser}
\affiliation{Technical University of Munich, TUM School of Natural Sciences, Garching 85748, Germany}
\affiliation{Walther-Mei{\ss}ner-Institut, Bayerische Akademie der Wissenschaften, Garching 85748, Germany}

\author{Abhishek Agarwal}
\affiliation{National Physical Laboratory, Teddington, TW11 0LW, United Kingdom}

\author{Ivan Rungger}
\affiliation{National Physical Laboratory, Teddington, TW11 0LW, United Kingdom}
\affiliation{Department of Computer Science, Royal Holloway, University of London, Egham, TW20 0EX, UK}

\author{Santana Lujan}
\affiliation{German Aerospace Center (DLR), Weßling, Germany}

\author{Max Werninghaus}
\affiliation{Walther-Mei{\ss}ner-Institut, Bayerische Akademie der Wissenschaften, Garching 85748, Germany}

\author{Stefan Filipp}
\affiliation{Technical University of Munich, TUM School of Natural Sciences, Garching 85748, Germany}
\affiliation{Walther-Mei{\ss}ner-Institut, Bayerische Akademie der Wissenschaften, Garching 85748, Germany}

\author{L\'eo Van Damme}
\email{leo.van-damme@tum.de} 
\affiliation{Technical University of Munich, TUM School of Natural Sciences, Garching 85748, Germany}
\affiliation{Munich Center for Quantum Science and Technology (MCQST), Schellingstra{\ss}e 4, M\"{u}nchen 80799, Germany}

\author{Steffen J. Glaser}
\email{glaser@tum.de}
\affiliation{Technical University of Munich, TUM School of Natural Sciences, Garching 85748, Germany}
\affiliation{Munich Center for Quantum Science and Technology (MCQST), Schellingstra{\ss}e 4, M\"{u}nchen 80799, Germany}

\date{\today}

\begin{abstract}
Dynamical decoupling (DD) is a widely used and resource‑efficient technique for error suppression, but conventional DD relies on periodically repeating a short pulse block to refocus the qubit state during idle periods. Imperfections in this block cause residual errors to accumulate, ultimately degrading state recovery over long idle times. Here, we introduce a universal optimal tracking approach that extends the original tracking concept to a fully state‑independent setting for designing DD sequences. By monitoring the qubit’s evolution at predefined waypoints during optimization, the method dynamically compensates residual errors while preserving regular refocusing. Experimental demonstrations on a superconducting‑qubit platform confirm the suppression of error accumulation under static control imperfections, in agreement with numerical predictions. Complementary simulations further show that optimal‑tracking‑based sequences maintain strong performance under time‑dependent noise. These results establish optimal tracking as a practical and hardware‑agnostic approach to designing short, robust DD sequences suitable for noisy quantum devices.
\end{abstract}

\maketitle
\section{Introduction} \label{sec:intro}
Quantum systems are inherently fragile and highly susceptible to environmental noise, leading to decoherence and posing a significant challenge for their reliable operation~\cite{shor1995scheme, preskill2018quantum, krantz2019quantum, google2025quantum}. Dynamical decoupling (DD) is among the earliest and most widely adopted techniques for suppressing errors in quantum systems. Building on the concept of spin echoes developed by Erwin Hahn in 1950~\cite{hahn1950spin}, the elimination of unwanted interactions has been the subject of various decoupling techniques~\cite{carr1954effects, meiboom1958modified,Levitt1982Supercycles, Gullion1990, souza2011robust}. 
DD has been successfully implemented across a variety of quantum platforms, including superconducting qubits~\cite{bylander2011noise,ezzell2023dynamical,gustavsson2012dynamical,google2025quantum, han2025protecting}, trapped ions~\cite{biercuk2009optimized, biercuk2009experimental}, neutral atoms~\cite{sagi2010process}, photonic qubits~\cite{damodarakurup2009experimental}, and quantum sensing~\cite{Degen17,wang2012comparison}. The core principle of DD is to preserve quantum coherence by applying precisely timed control pulses that average out unwanted system-environment interactions~\cite{meiboom1958modified, maudsley1986modified, genov2017arbitrarily}. In today’s noisy quantum computers, it has become a powerful tool for enhancing the fidelity of quantum algorithms~\cite{pokharel2024better, singkanipa2025demonstration, pokharel2023demonstration, tong2025empirical, rahman2024learning, google2025quantum}, and is now routinely incorporated into circuit compilation routines for various hardware platforms~\cite{QiskitCommunity2017}.\\

Over the years, numerous DD sequences have been developed, each offering different levels of protection against specific types of noise, as reviewed in the studies~\cite{souza2012robust, suter2016colloquium, ezzell2023dynamical}. For instance, the CPMG sequence is known not to be universal~\cite{ezzell2023dynamical,souza2011robust}, meaning it does not protect all quantum states equally. In contrast, sequences such as XY4 are universal and also robust up to some order against pulse imperfection errors such as drive frequency deviations (detuning) and pulse-amplitude deviations (over- or under-rotations)~\cite{ezzell2023dynamical}. Basic DD sequences simply repeat the same pulse or block of pulses throughout the decoupling sequence. However, in practical scenarios where the pulse block does not perfectly refocus the qubit state (e.g., due to imperfect pulses, the net propagator deviates away from identity), residual errors can accumulate across successive repetitions, leading to poor state recovery at the end of the idle time. Consequently, enhancing the robustness of these sequences remains a key objective in the design of dynamical decoupling protocols.~\cite{genov2017arbitrarily}.

To address this challenge, several strategies have been proposed, such as using composite pulses~\cite{levitt1979nmr,levitt1986composite,freeman1998spin}, which give rise to sequences like KDD~\cite{souza2011robust}, or by concatenating existing sequences, as in CDD~\cite{khodjasteh2005fault,khodjasteh2007performance}. However, these approaches often lead to an exponential increase in the number of required pulses to achieve certain robustness~\cite{souza2012robust, ezzell2023dynamical}. The universal robust ($\text{UR}$) family of DD sequences addresses this challenge by providing robustness to a broader range of pulse imperfections with only $N$ pulses per block~\cite{genov2017arbitrarily}. Notably, low-order UR sequences reduce to well-established DD sequences (e.g., for $N = 2$, $\text{UR}2$ is identical to $\text{CPMG}$ and for $N = 4$, $\text{UR}4$ is identical to $\text{XY4}$), indicating that their robustness is similarly limited for small $N$. We note that the term ‘universal robust’ (UR) refers to robustness against a broad class of pulse imperfections, and should not be confused with ‘universal rotation’ pulses~\cite{levitt1986composite, Kobzar2012, skinner2012new}, which denote pulses that produce the same rotation angle regardless of resonance offset or rf-amplitude inhomogeneity. Throughout this paper, UR refers exclusively to universal robust DD sequences. A complementary approach exploits supercycle constructions, where short four-pulse blocks such as $\text{XY}4$ or MLEV-type sequences are combined with cyclic phase shifts and sign reversals (e.g., $\text{XY}64$ and related schemes) to enhance robustness while retaining a four-pulse structure~\cite{Levitt1982Supercycles,Gullion1990,levitt1981compensation}. A related strategy is to randomize the order of the control operations from block to block, which slows the accumulation of residual averaging errors in long sequences~\cite{santos_RDD, santos_advantages_2008}. In practice, however, shorter base sequences are often essential in large‑scale quantum circuits, where idle windows of varying duration arise between gate operations. Such short DD blocks can be inserted into these temporal gaps without disrupting the surrounding control operations. Moreover, short sequences enable more frequent refocusing, i.e., periodically returning the qubit state close to its initial state, which improves resilience to time-dependent noise, a major source of decoherence arising from fluctuating system-environment coupling, for example, due to two-level fluctuators~\cite{Agarwal2025fast, Martinis2005}. These considerations highlight the need for DD sequences that are both short and intrinsically robust, not only against static detuning errors and pulse imperfections but also against time-dependent noise, so that they continue to average out unwanted dynamics even when the environment is not stationary.\\
\begin{figure}[!]
    \centering
    \includegraphics[width=1\linewidth]{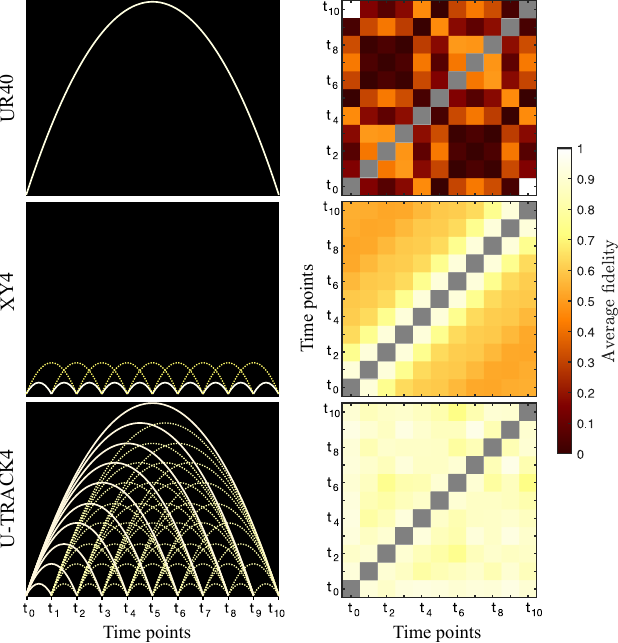}
    \caption{Comparison of the design principles of two conventional DD sequences (UR40, top row, and XY4, middle row) with the novel tracking-based approach introduced in this paper (U-TRACK4, bottom row). On the left, the white arcs represent propagators $U({t_{m},t_{n}})$ designed to approximate the identity operator $\mathbb{I}$ from time point $t_m$ to time point $t_n$ for $0\leq m<n\leq10$. Yellow dashed lines indicate propagators that, by direct transitivity, also approximate the identity operator. On the right, fidelities measuring the proximity of $U({t_{m},t_{n}})$ to $\mathbb{I}$, averaged over a range of frequency detunings and pulse-amplitude deviations, are shown. Trivial diagonal entries with $U({t_{n},t_{n}}) = \mathbb{I}$ are grayed out. The vertical axis of the left panels is omitted as it carries no physical meaning; only the extent of the arcs is relevant.} 
    \label{fig:FigIntro}
\end{figure}

In this work, we introduce a new family of dynamical decoupling sequences, denoted as U-TRACK, based on the idea of \textit{Universal optimal TRACKing}. The key idea of tracking is to enforce, during optimization, that the qubit’s evolution approximates the identity not only at the end of the sequence but also at intermediate checkpoints, i.e., after each pulse block, thereby preventing error accumulation across successive cycles. The optimal tracking framework was originally developed to efficiently decouple two heteronuclear spins in NMR~\cite{neves2009heteronuclear,schilling2014next}, where a GRAPE-based algorithm~\cite{Khaneja2005} was used to optimize (non-periodic) decoupling sequences that transfer a defined initial state to a desired target state not only at the end of the sequence (the final time) but also at specified tracking points during the sequence. Here, we generalize this concept to a universal control setting, i.e., one that is independent of the input state, to design dynamical decoupling sequences.  This approach enables a design principle that fundamentally differs from conventional DD approaches, as illustrated in Fig.~\ref{fig:FigIntro}.\\ 

This approach enables the design of a DD sequence that provides dynamic compensation to suppress error accumulation. Unlike basic DD sequences, which rely on repeating identical pulse blocks and thus allow systematic errors to accumulate, the resulting U-TRACK sequences introduce corrections at each block, yielding non-periodic pulse structures. Consequently, robustness against parameter deviations, such as detuning and amplitude errors, remains stable even as the number of pulse blocks increases. This effect can be directly observed in the fidelity plots on the right of Fig.~\ref{fig:FigIntro}, where each entry $(t_m, t_n)$ shows the average process fidelity of the propagator $U(t_m, t_n)$ with respect to the identity, averaged over a range of detuning and amplitude errors. U-TRACK4 maintains high fidelity at all intermediate times, reflected by bright entries throughout the matrix, while XY4 shows gradual degradation visible as darkening entries away from the diagonal. For longer sequences such as UR40~\cite{genov2017arbitrarily}, where the entire sequence is optimized as a single identity gate between the initial and final times, fidelity is high at the endpoint by design, but intermediate-time fidelity is unconstrained and typically poor, as visible in the irregular fidelity matrix of UR40 in Fig.~\ref{fig:FigIntro}. As a result, any sudden change in the dynamics not covered by the model, such as a mid-sequence parameter jump, will degrade fidelity even at the endpoint, as explicitly discussed later. We note that this block‑by‑block tracking structure is closely related to supercycle‑based DD (e.g., XY64), but U‑TRACK generalizes these constructions by optimizing arbitrary blocks and robustness regions. Moreover, the optimized U-TRACK sequences do not simply recover known supercycles: the optimization naturally converges to a structure — which we term XUR — that mixes MLEV-type and XY-type blocks within a single sequence, a combination not previously considered in the literature~\cite{Levitt1982Supercycles, Freeman1982, Gullion1990}.\\

Our optimal tracking approach optimizes the pulse phases of each block using a combination of GRAPE-based gradient optimization and a genetic algorithm. The GRAPE component efficiently computes local gradients, while the genetic algorithm performs a structured global search across the pulse space, helping to avoid suboptimal local minima and increasing the likelihood of finding globally optimal solutions. The method supports an arbitrary number of blocks, $M$, and any even number of pulses, $N$, per block. To highlight the novelty and practical advantages of the tracking approach, we focus on the $N=4$ sequence U-TRACK4 and investigate its robustness to parameter deviations both numerically and experimentally on superconducting-qubit hardware, comparing its performance against XY4 and UR40. The experimentally obtained robustness profiles show excellent agreement with the numerical results, as detailed in Sec.~\ref{Sec.Robust_para}.\\

In Sec.~\ref{Sec.Rob_time}, we examine the robustness of different DD protocols to time-dependent detuning noise and evaluate their average fidelity across an ensemble of random-telegraph-noise realizations, enabling a direct comparison of sequence performance as a function of switching rate. In Sec.~\ref{sec:structure}, we analyze the structure of the optimized sequences and show that they converge to a novel XUR form (a mixture of MLEV-type and XY-type blocks) and discuss how this structure can serve as an effective warm-start for the optimization, which in practice tends to reduce the optimization time. The optimization framework used in this work is available as a Julia package, UTrack.jl~\cite{malvetti2024utrack}, and ready‑to‑use optimized sequences for various lengths $N$ are provided in the package repository. The resulting sequences are hardware‑agnostic; experimental validation in this study was performed on a superconducting quantum device.
    
\section{Optimal tracking} \label{sec:methods}
\begin{figure*}[t]
    \includegraphics[width=1\linewidth]{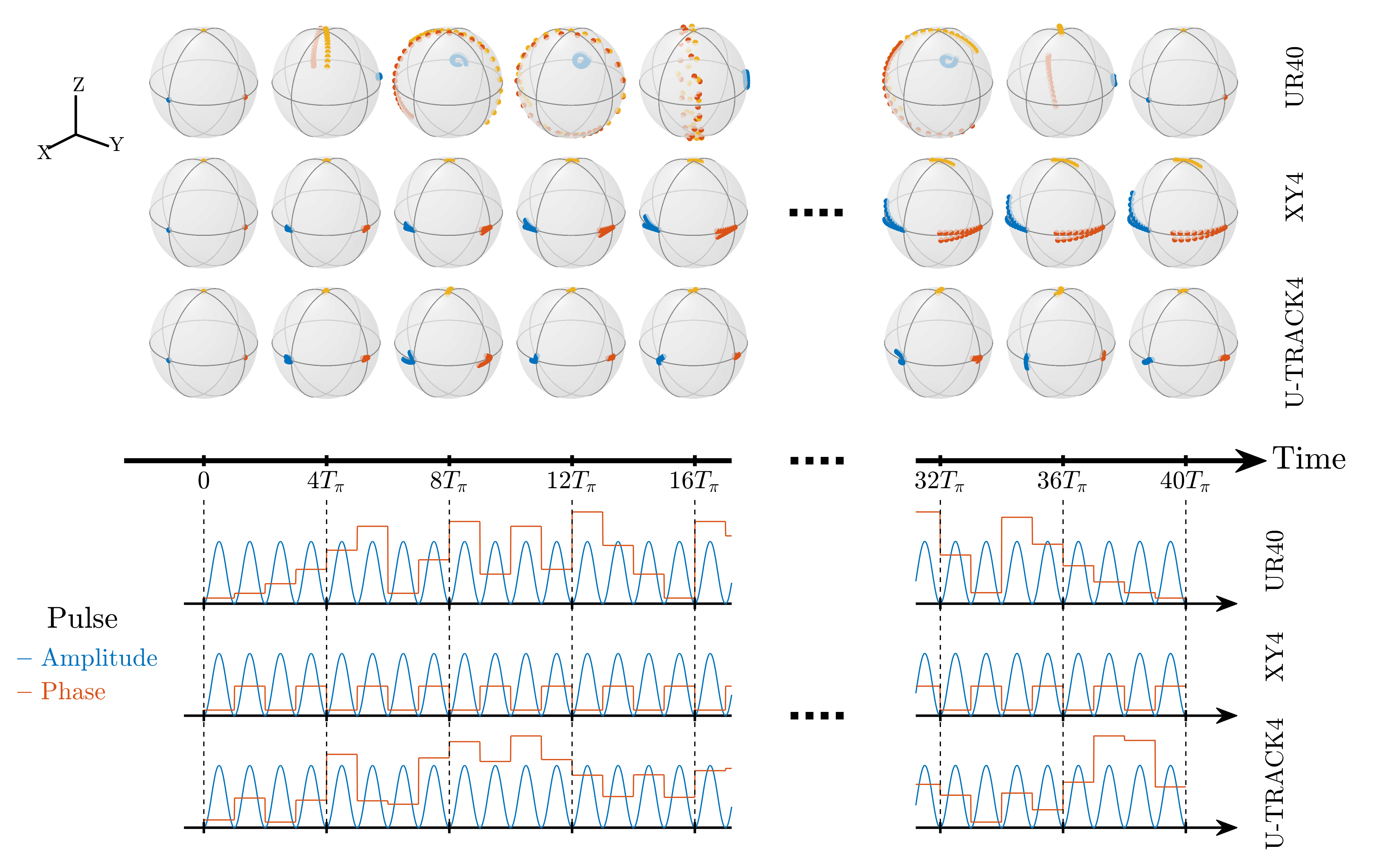}
    \caption{Effect of three dynamical decoupling sequences on qubit dynamics, shown on the Bloch sphere (top) and as pulse waveforms (bottom). Each row corresponds to one sequence (UR40, XY4, U-TRACK4); each sphere is a snapshot in time, progressing from $t=0$ to $t=40\,T_{\pi}$, where $T_{\pi}$ is the duration of one $\pi$-pulse. Points mark qubit states initialized on $X$ (blue), $Y$ (red), and $Z$ (yellow). Each evolves under a different detuning strength; a tight cluster therefore indicates resilience to detuning errors, while a spread or displaced set of points indicates sensitivity. Ideally, every point returns to its starting position by $T = 40\, T_{\pi}$. UR40 achieves a very robust state recovery at $T$ but not at intermediate times. XY4 recovers states after every block of four $\pi$-pulses, but small residual errors accumulate across repetitions, visible as a gradual drift of the point clusters. U-TRACK4, derived via optimal tracking control, matches XY4's block-by-block recovery while suppressing error accumulation, keeping the clusters tightly concentrated throughout the entire sequence.}
    \label{FigBloch}
\end{figure*}
This section introduces the optimal tracking approach and compares it with existing DD methods. We first provide a qualitative overview of how the three approaches — single-block optimization, periodic repetition, and optimal tracking — differ in their design philosophies and practical performance. To make these differences concrete, we intentionally introduce experimental imperfections, such as static detuning and pulse-amplitude errors, and compare the resulting fidelities both numerically and experimentally. Full technical details on the U-TRACK optimization algorithm, the experimental implementation on a transmon-based quantum processor, and the transmon qubit simulations are provided in Appendices~\ref{app:optimization},~\ref{App:WMI_experiment}, and~\ref{app:simulations} respectively.
\subsection{Dynamical decoupling schemes}
\label{Sec.DD_schemes}
A dynamical decoupling (DD) sequence protects a qubit’s state during idle periods — times when the qubit is not actively used — by applying a series of control pulses that counteract noise and environmental disturbances. Without DD, fluctuations in the qubit’s environment can cause its state to drift, introducing errors into quantum-circuit outcomes. The two dominant error types are frequency detuning — which induces unwanted rotations about the $z$-axis — and pulse-amplitude deviations — which lead to over- or under-rotations around a transverse axis. On transmon-based devices, these errors may arise from imperfect calibration~\cite{Sheldon2016,Wright_SC}, crosstalk~\cite{Tripathi2022}, interactions with two-level systems (TLSs)~\cite{Agarwal2024,Martinis2005}, flux noise~\cite{Kumar2016} or 
pulse distortions~\cite{Sheldon2016, Hyyppa2024}. A DD sequence is designed so that, despite these imperfections, the qubit’s net evolution during an idle time $T$ is effectively the identity, i.e., $U(T) = \mathbb{I}$, where $U(T)$ denotes the qubit’s evolution operator.

We compare three dynamical decoupling techniques:
\begin{enumerate}
\renewcommand{\labelenumi}{(\roman{enumi})}
\item A single sequence of pulses spanning the entire idle period, such as UR40.
\item Repeated application of a shorter base sequence, such as XY4.
\item Optimal tracking (U-TRACK): a long sequence designed to periodically refocus the qubit state while suppressing error accumulation across blocks.
\end{enumerate}

Here, U-TRACK optimizes a sequence of gates, namely $\pi$-rotations with optimized phases, to enable a direct comparison with existing DD methods. The method is, however, not limited to $\pi$-gates: general pulse-phase shaping could provide more degrees of freedom, which we leave for future work.\\

Approach~(i) enables the design of sequences that are robust against large deviations in the Hamiltonian parameters, since the additional pulse energy available during longer idle periods allows for greater error compensation. In ref.~\cite{genov2017arbitrarily}, analytical formulas are provided to construct sequences that can span arbitrarily long idle times. These sequences, referred to as the universal robust (UR) family, are composed of multiple $\pi$-pulses with specific phases, and become increasingly robust against detuning and amplitude errors as the idle period — and correspondingly the number of $\pi$-pulses — grows, assuming these errors remain constant in time. However, while this assumption generally holds for a few gates acting on a qubit, additional events during longer idle periods may alter the Hamiltonian, reducing the sequence's effectiveness.\\

In contrast, the self-repeating structure of approach~(ii) enables periodic refocusing of the qubit state. This repeated refocusing reduces sensitivity to time-dependent parameter fluctuations compared to approach~(i). However, because it relies on repeating a short sequence with limited intrinsic robustness, residual errors accumulate across successive cycles, reducing robustness to parameter deviations during long idle periods.\\

Approach~(iii) combines the strengths of approaches~(i) and~(ii) while overcoming their respective limitations. These sequences are referred to as the U-TRACK family. As in approach~(i), a U-TRACK sequence is optimized over an arbitrarily long time, but it also refocuses the qubit state regularly, as in approach~(ii). This optimization prevents error accumulation across cycles, yielding a sequence that remains effective in the presence of time-dependent fluctuations and robust against large parameter deviations, even over long idle periods.\\

One can also consider a fourth approach: repeating short DD sequences with small modifications between blocks, such as MLEV sequences or generalizations of XY4~\cite{Levitt1982Supercycles,Freeman1982,Gullion1990,lancaster2025}. For example, XY64 combines several variations of XY4, where pulses are ordered as XYXY or (-X)(-Y)(-X)(-Y), and each can be cyclically shifted to YXYX or (-Y)(-X)(-Y)(-X). By construction, these sequences also regularly refocus the qubit state. These sequences share structural properties with optimal tracking, as discussed in Sec.~\ref{sec:structure}. Nevertheless, this technique is limited to blocks of four $\pi$-pulses, whereas U-TRACK or UR pulses can support any even block length.\\

To illustrate the three approaches, we consider three DD schemes, namely UR40, XY4, and U-TRACK4, which correspond to approaches (i), (ii), and (iii), respectively. The sequences are illustrated in Fig.~\ref{FigBloch}. UR40 consists of 40 $\pi$-pulses with phases computed using the analytical formula in Ref.~\cite{genov2017arbitrarily}. XY4 is a sequence of four $\pi$-pulses with phases $\big{(}0, \frac{\pi}{2}, 0, \frac{\pi}{2}\big{)}$, repeated 10 times to cover the same idle period, and U-TRACK4 is a sequence of 40 $\pi$-pulses with phases optimized using optimal tracking, taken as the first 40 pulses of a sequence optimized over 100 blocks (Appendix~\ref{app:optimization}). Figure~\ref{FigBloch} illustrates the resulting qubit dynamics for states $X$, $Y$, and $Z$, under experimentally realistic detuning errors. We observe that UR40 recovers the state very accurately at the end of the full sequence, but some states are widely spread over the Bloch sphere at intermediate times. In contrast, XY4 recovers the states after each cycle of four pulses; however, because it is not perfectly recovered after a single cycle, the states slowly drift over successive cycles. The U-TRACK4 sequence combines the advantages of both techniques: it refocuses the states after every block of four $\pi$-pulses while avoiding error accumulation. We note that supercycle-based sequences such as XY64 exhibit behavior similar to U-TRACK4 for four-pulse blocks, as discussed in Appendix~\ref{app:XY64}.

\subsection{Robustness to parameter deviations}
\label{Sec.Robust_para}
\begin{figure*}
    \centering
    \includegraphics[width=0.65\linewidth]{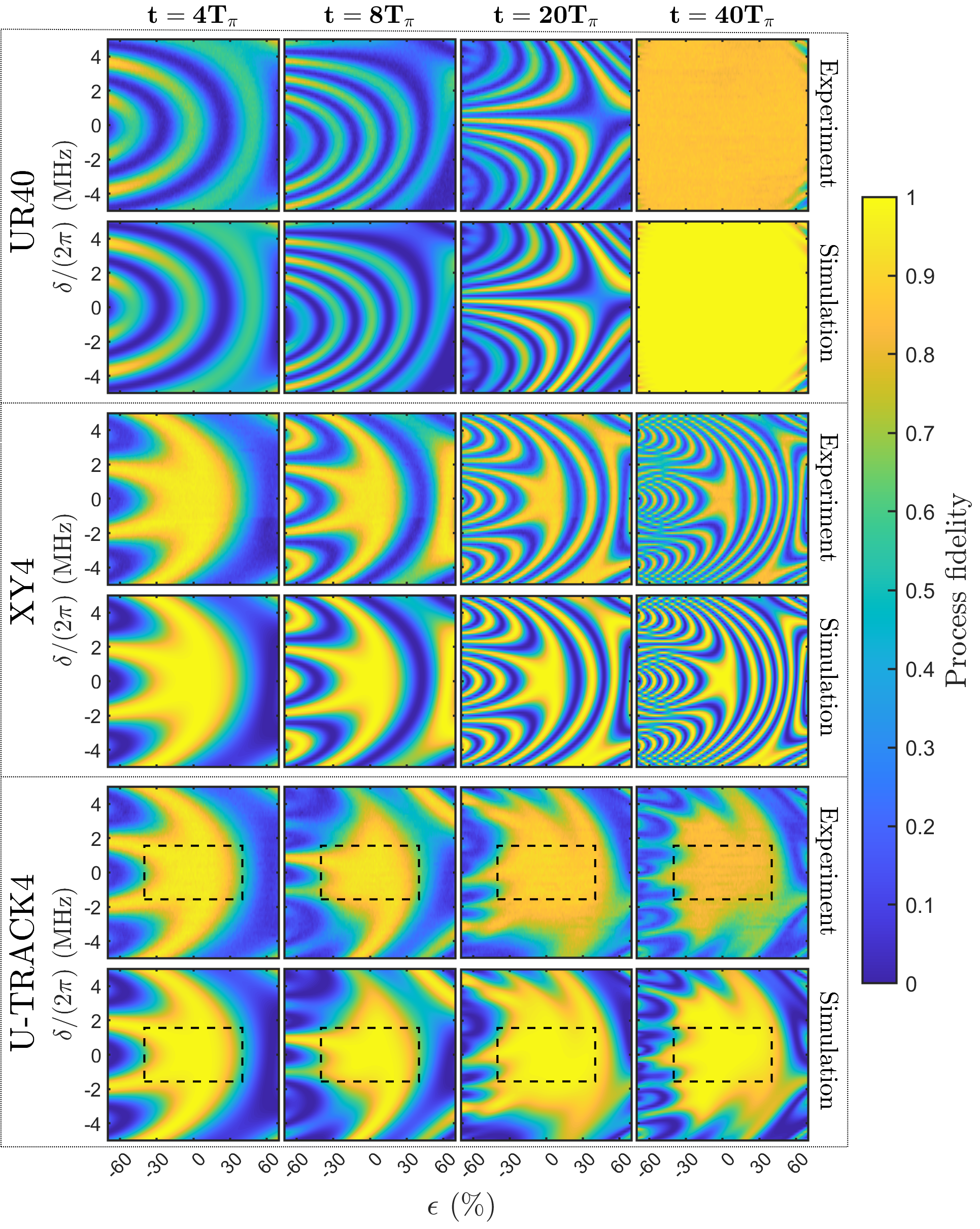}
    \caption{Process fidelities of different dynamical decoupling sequences as functions of pulse-amplitude error $\epsilon$ and frequency detuning $\delta$, which are intentionally engineered in the quantum device. For each sequence, the top row shows experimental measurements, while the bottom row shows the corresponding simulation results. Each column corresponds to the fidelity obtained after $4$, $8$, $20$, and $40$ $\pi$-pulses, respectively, with each $\pi$-pulse having a duration of $T_{\pi}=128$~ns. The dotted rectangle shows the parameter ranges over which the U-TRACK4 sequence was optimized with assigned weights in the central region, as described in Appendix~\ref{app:optimization}. Technical details of the experimental implementation and simulations are reported in Appendices~\ref{App:WMI_experiment} and~\ref{app:simulations}, respectively.}
    \label{fig:FidelityMaps}
\end{figure*}

The differences in robustness between the three dynamical decoupling approaches are illustrated in Fig.~\ref{fig:FidelityMaps}. 
In this figure, amplitude deviations $\epsilon$ and static detuning $\delta$ were intentionally applied on the quantum device to mimic experimental imperfections, and the fidelities of each DD sequence were computed from quantum process tomography measurements~\cite{NielsenChuangBook} for every $(\epsilon, \delta)$ pair as described in Appendix~\ref{App:WMI_experiment}. Quantum process tomography quantifies the fidelity of the identity gate acting on a particular qubit, offering a general assessment of the DD sequence’s performance for any input state. \\

In this example, we consider an idle period equal to the duration of 40~$\pi$-pulses of 128~ns each without any inter-pulse delays, that is, a total idle time of $T = 5.12~\mu\text{s}$. One can see that UR40 delivers good performance over a very broad range of amplitude errors $\epsilon$ and frequency detunings $\delta$, but only once the full sequence has been completed (i.e., at $t = 40\, T_{\pi}$). At intermediate times, its fidelity landscape appears irregular, since the pulse is not designed to refocus at shorter timescales. In contrast, the fidelity of XY4 is only high over a reasonably broad range of parameters after one or two XY4 cycles, but this range gradually shrinks as more cycles are applied, reflecting the accumulation of residual errors. As a result, by the end of the full idle time, the sequence exhibits limited robustness (concentrated around the center). U-TRACK4, by comparison, remains robust even after the full sequence of $40$~$\pi$-pulses, while still benefiting from regular refocusing. The only visible difference between experiment and simulation is a global loss of fidelity due to decoherence, which the simulations do not include. We also observed similar robustness behavior when a delay is introduced between the pulses.  

\subsection{Robustness to time-dependent noise}
\label{Sec.Rob_time}
\begin{figure*}[t]
    \centering
    \includegraphics[width=0.8\linewidth]{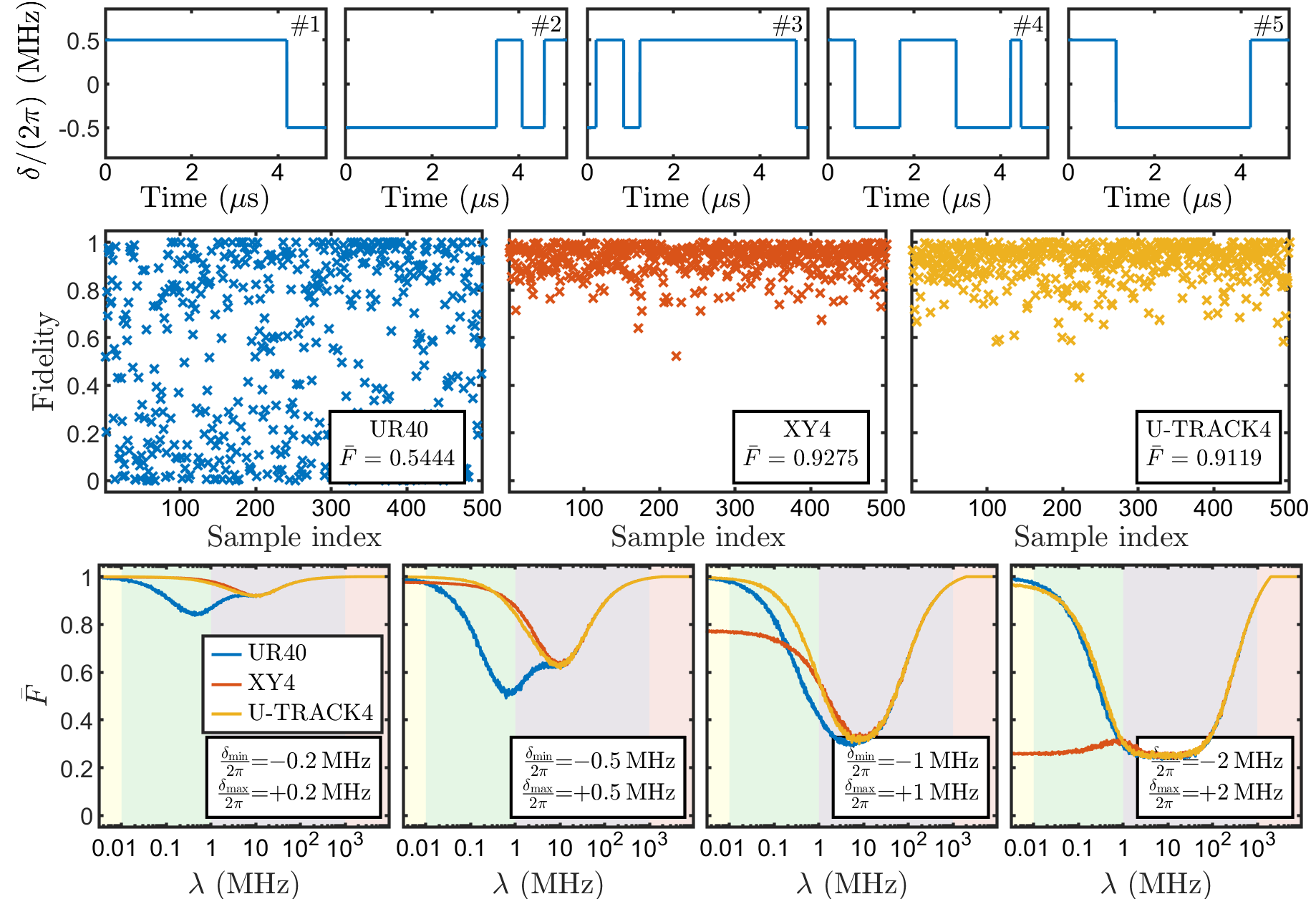}
    \caption{\emph{First row:} Five representative samples of detuning versus time under random telegraph noise (RTN) generated with a switching rate $\lambda=0.5$~MHz, and detuning levels of $\pm 0.5$~MHz, illustrating the stochastic variability of noise. \emph{Second row:} Simulated sequence fidelity at the idle time ($T=5.12~\mu$s, corresponding to 40~$\pi$-pulses) computed for 500 samples ($\lambda = 0.5$~MHz, $\delta(t)/(2\pi)=\pm 0.5$~MHz). Results are shown for the three dynamical decoupling sequences indicated in the legend. $\bar{F}$ denotes the average fidelity over the 500 samples.
    \emph{Third row:} Average fidelity over 2,000 samples as a function of $\lambda$, shown for different detuning levels as indicated in the legend. The different-colored strips in the plots indicate different frequency regimes, as explained in the main text.} \label{fig:TimeDependentNoise}
\end{figure*}

The fidelity profile in Fig.~\ref{fig:FidelityMaps} illustrates the robustness of each sequence against static deviations. It does not capture the response to time-dependent fluctuations in the Hamiltonian parameters. Such fluctuations can occur during the execution of quantum circuits due to mechanisms discussed earlier~\cite{Hyyppa2024,Tripathi2022,Agarwal2024,Martinis2005, Agarwal2025fast}. In particular, two-level fluctuators~\cite{Agarwal2025fast} or charge parity switches~\cite{Tennant2022, Agarwal2025fast}, can cause the detuning to switch between two values, $\delta_{\min}$ and $\delta_{\max}$. These fluctuations can be modeled as \emph{random telegraph noise} (RTN)~\cite{Paladino2014,Cywinski2008}, a type of stochastic noise in which a parameter --- in this case, the detuning --- abruptly switches between two values at random times. The characteristics of the noise can be quantified by the switching rate, $\lambda$ (in Hz), which represents the average number of times the detuning $\delta(t)$ switches per second. As a concrete example, for $T = 5.12~\mu$s and a switching rate of $\lambda = 1$~MHz, the expected number of switches during the idle period is $\lambda T \approx 5$, meaning the detuning switches approximately five times on average.\\

To evaluate the performance of a dynamical decoupling sequence under this noise, we simulate the qubit’s evolution for individual realizations of the detuning $\delta(t)$, using the same Gaussian $\pi$-pulses as in the experiment (see Appendix~\ref{app:simulations}). For a specified value of $\lambda$, we generate a large ensemble of random detuning profiles---typically 2,000 samples. By computing the sequence fidelity for each sample, we can statistically quantify the sequence’s effectiveness in suppressing the effects of RTN.\\

The first row of Fig.~\ref{fig:TimeDependentNoise} displays five representative samples of time-dependent detuning to illustrate the variability of detuning with time. The second row of Fig.~\ref{fig:TimeDependentNoise} displays the sequence fidelities for 500 such samples. The results show that UR40 performs notably worse than XY4 and U-TRACK4, with an average fidelity of $\bar{F}\simeq 0.5$ versus $\simeq 0.9$ for the other two sequences. Moreover, the fidelity of individual UR40 realizations spans nearly the full range from 0 to 1, reflecting high sample-to-sample variability, whereas XY4 and U-TRACK4 remain consistently close to unity across almost all realizations. This discrepancy arises from structural differences in the pulse sequences. Specifically, UR40 does not refocus the state until the end of the full sequence, whereas both XY4 and U-TRACK4 refocus the state after every block of four $\pi$-pulses (every $512~\mathrm{ns}$). Consequently, XY4 and U-TRACK4 exhibit reduced sensitivity to detuning fluctuations. However, this example is limited to $\lambda=0.5$~MHz (slow fluctuations); the behavior across a broader range of noise regimes is examined in the following.\\

The last row of Fig.~\ref{fig:TimeDependentNoise} gives a deeper understanding of how these sequences behave in more general scenarios. It shows the average fidelity $\bar{F}$ over 2,000 samples as a function of $\lambda$ for various detuning amplitudes. Generally speaking, U-TRACK4 performs comparably to, or better than, the best of XY4 and UR40 for $\lambda\lesssim 10$~MHz, while all three sequences perform similarly for $\lambda \gtrsim 10$~MHz. Looking in more detail, four distinct regimes can be distinguished:  
\begin{itemize}
    \item High-frequency noise: $\lambda \gtrsim 10^3$~MHz (pink area).
    \item Medium-frequency noise: $\lambda \sim 1$–$10^3$~MHz (purple area).
    \item Low-frequency noise: $\lambda \sim 0.01$–$1$~MHz (green area).  
    \item Static detuning: $\lambda< 0.01$~MHz (yellow area). The switching rate is so low that a switch during the idle time is unlikely. Only two outcomes are therefore possible: $\delta = \delta_{\max}$ or $\delta = \delta_{\min}$.  
\end{itemize}

For high-frequency noise (pink area), all sequences perform equally well (near $\bar{F} = 1$). In this regime, the detuning fluctuates so rapidly that its effect averages out to zero, since each sample switches between $\delta_{\max}$ and $-\delta_{\max}$.\\

For medium-frequency noise (purple area), all sequences perform relatively poorly. In this regime, the regular refocusing of XY4 and U-TRACK4 does not provide a clear advantage over UR40 in compensating for the fluctuations. \\

For low-frequency noise (green area), the differences between the three sequences become clear. In the first two panels of the last row, XY4 and U-TRACK4 maintain significantly higher average fidelity than UR40. In the second panel, for example ($\delta/(2\pi)=\pm 0.5$ MHz), UR40 drops to roughly 0.6 around $\lambda \sim 0.2$~MHz, while XY4 and U-TRACK4 remain above 0.9. In the right-most panel, XY4 performs noticeably worse, with a fidelity around 0.2 for $\lambda \lesssim 0.1$~MHz. This behavior occurs because, at such low $\lambda$, the detuning is effectively quasi-static, producing an effect similar to that discussed below.\\

For static detuning (yellow area), all sequences perform similarly in the first two panels, where $\delta_{\max}/(2\pi) \leq 0.5~\mathrm{MHz}$. In this regime, the detuning is constant and relatively small, so all sequences compensate for its effect. Indeed, as shown in the last column of Fig.~\ref{fig:FidelityMaps}, a detuning of $\pm 0.5~\mathrm{MHz}$ lies well within the high-fidelity region for all sequences. In contrast, a detuning of $\delta/(2\pi) = 1~\mathrm{MHz}$ is at the edge of this high-fidelity region for XY4. This explains why XY4 performs relatively poorly in the last two panels of Fig.~\ref{fig:TimeDependentNoise} near $\lambda\sim 0.01$~MHz.\\

In addition, we observe that for low-frequency noise, XY4 sometimes achieves slightly higher fidelity than U-TRACK4. This is particularly visible in the second panel of the last row, for $\lambda \sim 1$~MHz. The origin of this feature is not entirely clear, but it may be related to the specific phase pattern of the sequences, which can have subtle effects. Nevertheless, the difference is very small and is not expected to have any significant impact in practice. \\

\begin{figure}[h!]
    \centering
    \includegraphics[width=1\linewidth]{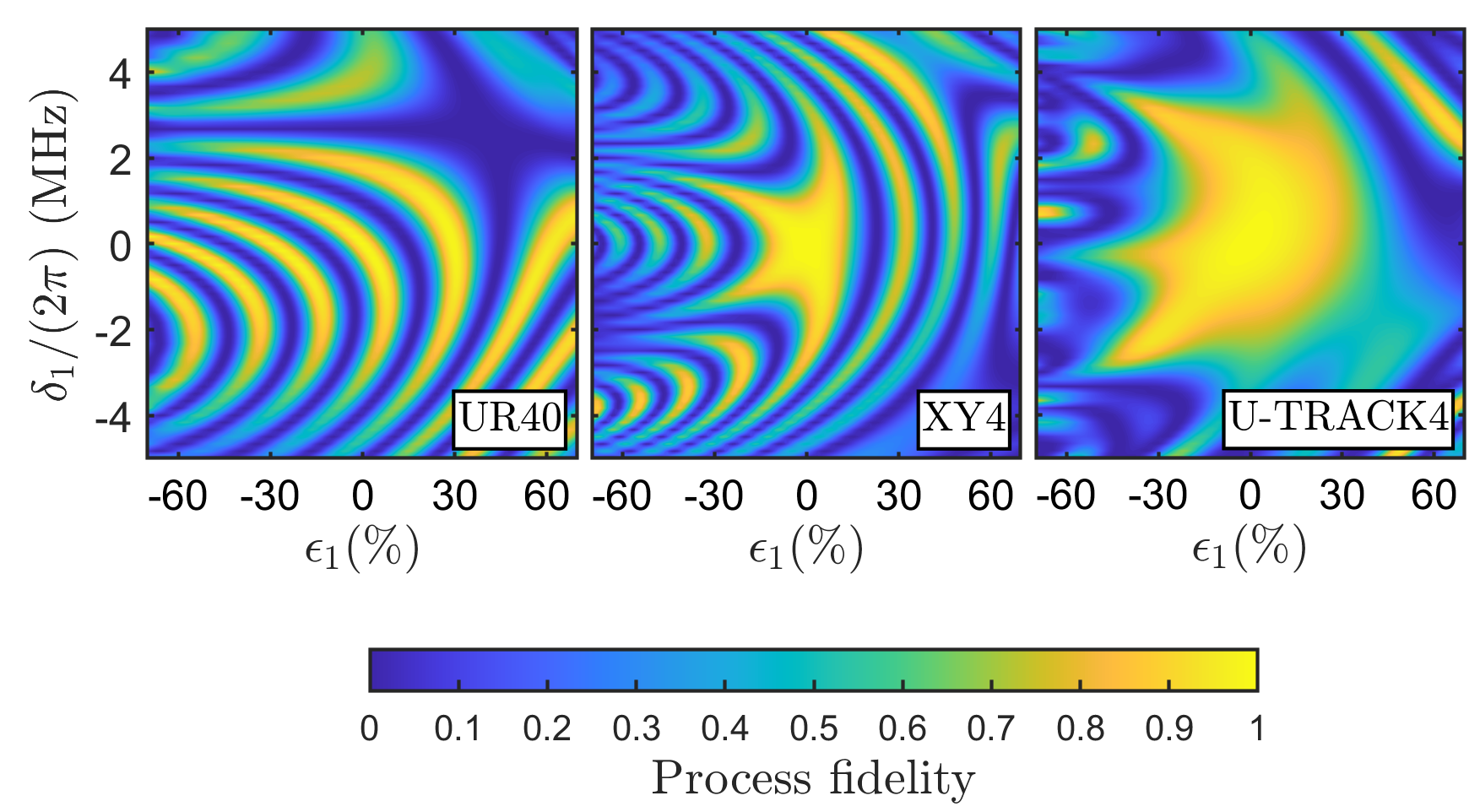}
    \caption{Simulated fidelity of the three sequences at the end of the idle time as a function of $(\epsilon_1, \delta_1)$, for a system starting at $(\epsilon_0, \delta_0) = (0, 0)$ and jumping to $(\epsilon_1, \delta_1)$ at $t = T/3 = 1.707~\mu$s, simulated with the Gaussian $\pi$-pulses described in Appendix~\ref{app:simulations}. Compared to the static case (last column of Fig.~\ref{fig:FidelityMaps}), XY4 and U-TRACK4 retain a centered, well-preserved high-fidelity region, whereas the high-fidelity region of UR40 is fragmented and shifted, revealing its sensitivity to mid-sequence parameter changes.}
    \label{fig:TDLandscape}
\end{figure}

The poor performance of UR40 in the low-frequency regime stems from the fragility of its fidelity profile to mid-sequence perturbations. To illustrate this, we consider a scenario in which the system parameters start at $(\epsilon_0, \delta_0) = (0, 0)$ and jump to $(\epsilon_1, \delta_1)$ at $t = T/3 = 1.707~\mu$s, analogous to the step-like profile shown in the top-left panel of Fig.\ref{fig:TimeDependentNoise}. Fig.~\ref{fig:TDLandscape} shows the simulated fidelity as a function of $(\epsilon_1, \delta_1)$ at the end of the idle time; this should be compared directly with the last column of Fig.~\ref{fig:FidelityMaps}, which shows the static case. The key observation is that for XY4 and U-TRACK4, the high-fidelity region remains centered and well-preserved under this perturbation, whereas for UR40, the high-fidelity region is not only reduced in area but also fragmented and shifted away from the center, revealing its sensitivity to mid-sequence parameter changes.

\section{Structure of the U-TRACK sequences} \label{sec:structure}
\begin{figure}[t]
\centering
\includegraphics[scale=0.8]{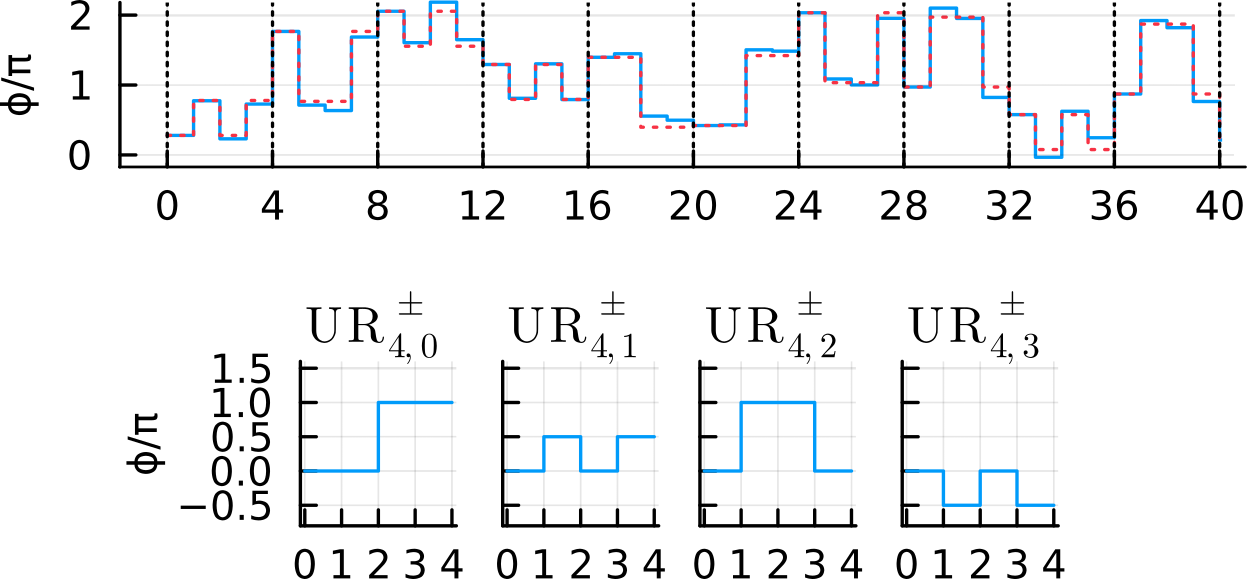}
\caption{Top: Optimized U-TRACK4 pulse phases for 10 blocks (solid blue). For comparison, the closest approximating XUR4 sequence (defined in the main text) is shown in red dotted. Bottom: The four distinct $\mathrm{UR}^\pm_{4,k}$ pulses ($k = 0, 1, 2, 3$). $\mathrm{UR}^\pm_{4,1}$ is equivalent to XY4 and $\mathrm{UR}^\pm_{4,0}$ equals the MLEV4 sequence~\cite{levitt1981compensation, levitt1982broadband, lancaster2025}. The remaining two cases are obtained by applying a cyclic phase shift and possibly adding a global phase offset. Details about the pulse optimization are provided in Appendix~\ref{app:optimization}.}
\label{fig:pulses}
\end{figure}

The structure of the optimized U-TRACK sequences provides insight into the underlying control mechanisms of robust DD and helps explain why the optimization reliably converges to high-performing solutions. In this section, we demonstrate that the U-TRACK sequences generated by our optimization algorithm exhibit simple, regular phase patterns and symmetries — even though no such structure was imposed as a constraint during optimization. Specifically, even when starting from random initial pulse phases, the optimization algorithm converges to pulse blocks whose phases follow a recognizable pattern closely related to known analytical DD sequences, as illustrated in Fig.~\ref{fig:pulses}. To understand why this structure emerges, we first analyze the symmetries of a single pulse block, i.e., a fixed sequence of $N$ $\pi$-pulses with phases $(\phi_1, \ldots, \phi_N)$ that is designed to approximate the identity over one refocusing period.

\begin{enumerate}[(i)]
\item Global phase shift: for any phase $\phi$, the sequences $(\phi_1,\ldots,\phi_N)$ and $(\phi_1+\phi,\ldots,\phi_N+\phi)$ yield the same fidelity, since the shift rotates the propagator about the z-axis (this rotation commutes with the detuning term), which leaves its trace invariant.

\item Cyclic permutation: The sequences $(\phi_1,\ldots,\phi_N)$ and $(\phi_2,\ldots,\phi_N,\phi_1)$ yield the same fidelity due to the cyclic property of the trace.

\item Reversing the pulse: The sequence $(\phi_N,\ldots,\phi_1)$ yields the same fidelity as $(\phi_1,\ldots,\phi_N)$ since $F(U^\dagger)=F(U)$ and $P(\phi,\epsilon,\delta)^\dagger=P(\phi+\pi,\epsilon,-\delta)$ (with $P$ as in Eq.~\eqref{eq:app_Pn}).
We use the fact that the detuning range is symmetric.

\item Negating all phases: The sequence $(-\phi_1,\ldots,-\phi_N)$ yields the same fidelity as $(\phi_1,\ldots,\phi_N)$ since $F(\Pi_y\, U\,\Pi_y)=F(U)$ and $\Pi_y P(\phi,\epsilon,\delta) \Pi_y=P(-\phi+\pi,\epsilon,-\delta)$ where $\Pi_y$ mirrors the $y$-axis.
Again, we use the fact  that the detuning range is symmetric.
\end{enumerate}

Note that while these symmetries hold exactly for a single block, they will only hold approximately across multiple blocks. The single-block case is well-understood through the UR pulses introduced in~\cite{genov2017arbitrarily}.
For any even $N$ greater than or equal to $4$, and any $k=0,\ldots,N-1$, one defines the sequence of phases $\phi_1,\ldots,\phi_N$ by
$$
\phi_i=\frac{(i-1)(i-2)}{2}\Phi_N  + (i-1)\frac{2\pi k}{N}
$$
where
$$
\Phi_{4m}=\pm\frac\pi{m},
\quad \Phi_{4m+2}=\pm\frac{2m\pi}{2m+1}\,.
$$
We denote this sequence by $\mathrm{UR}_{N,k}^\pm$. 
Note that in this definition $\phi_1=0$, but equivalent pulses can be obtained by adding an arbitrary overall phase.\\

The family of UR pulses is closed under the symmetry operations listed above:
If $\mathcal{S}((\phi_1,\ldots,\phi_N))=(\phi_2,\ldots,\phi_N,\phi_1)$ is the cyclic left shift, then 
$$ \mathcal{S}(\mathrm{UR}_{N,k}^\pm) = \frac{2\pi k}{N} + \mathrm{UR}_{N,k\pm s}^\pm $$
where $s=2$ if $N=0\mod 4$ and $s=\frac{N}2-1$ if $N=2\mod4$.
If $\mathcal{N}((\phi_1,\ldots,\phi_N))=(-\phi_1,\ldots,-\phi_N)$ flips all phases, then
$$ \mathcal{N}(\mathrm{UR}_{N,k}^\pm) = \mathrm{UR}_{N,N-k}^\mp\,. $$ 
Finally if $\mathcal{R}((\phi_1,\ldots,\phi_N))=(\phi_N,\ldots,\phi_1)$ reverses all the phases then
$$
\mathcal{R}(\mathrm{UR}_{N,k}^\pm) = \frac{2\pi (\pm s-k)}{N} + \mathrm{UR}_{N,\pm2s-k}^\pm\,.
$$

In the case of $\mathrm{UR}4$, the following diagram summarizes how the four pulses $\mathrm{UR}^\pm_{4,k}$ ($k = 0,1,2,3$) are related to each other via the symmetry operations $\mathcal{S}$ (cyclic shift), $\mathcal{N}$ (phase negation), and $\mathcal{R}$ (pulse reversal). This can be verified by inspecting the bottom panel of Fig.~\ref{fig:pulses}: for example, $\mathrm{UR}^\pm_{4,2}$ can be obtained from $\mathrm{UR}^\pm_{4,0}$ (MLEV4) by applying a cyclic shift to the phases, as reflected in the step pattern of the two panels.
\begin{equation*}
\begin{tikzcd}
{UR_{4,0}^\pm} \arrow[d, "\mathcal{S}"', bend right] \arrow["{\mathcal{N},\mathcal{R}}"', loop, distance=2em, in=125, out=55]  &  & {UR_{4,1}^\pm} \arrow[d, "{\mathcal{S},\mathcal{N},\mathcal{R}}"', bend right] \\
{UR_{4,2}^\pm} \arrow[u, "\mathcal{S}"', bend right] \arrow["{\mathcal{N},\mathcal{R}}"', loop, distance=2em, in=305, out=235] &  & {UR_{4,3}^\pm} \arrow[u, "{\mathcal{S},\mathcal{N},\mathcal{R}}"', bend right]
\end{tikzcd}
\end{equation*}

While UR blocks are typically repeated periodically to obtain longer sequences, periodically repeated pulses tend to accumulate errors over time, as discussed in Sec.~\ref{Sec.Robust_para}. More robust sequences can be obtained by combining different UR blocks of equal length — we call pulses of this general form XUR pulses. Indeed, the GRAPE-optimized U-TRACK pulses tend to take the form of XUR pulses (Fig.~\ref{fig:pulses}), i.e., a structured mix of different $\mathrm{UR}^\pm_{4,k}$ blocks rather than a periodic repetition of a single block. As a result, they outperform periodically repeated UR pulses in robustness (Fig.~\ref{fig:FidelityMaps}). Furthermore, initializing the GRAPE optimization from an XUR ansatz rather than a random initial guess leads to faster convergence, confirming that XUR pulses provide a physically motivated and effective starting point for numerical optimization.\\

This behavior closely resembles the fourth approach briefly mentioned in Sec.~\ref{Sec.DD_schemes}, which consists of applying short DD sequences with cyclic shifts or sign changes between blocks~\cite{Levitt1982Supercycles, Freeman1982, Gullion1990, lancaster2025}. Indeed, for this specific example, XY64 achieves similar performance to U-TRACK4, as shown in more detail in Appendix~\ref{app:XY64}. More generally, when using blocks of 4~$\pi$-pulses, U-TRACK reproduces the performance of XY64 or MLEV-type sequences; when using a single block spanning the full idle time, it recovers performance comparable to the UR family. Notably, the optimized U-TRACK sequences do not simply recover known supercycles: the XUR structure found by our optimization scheme mixes MLEV-type and XY-type blocks within a single sequence — a combination that, to our knowledge, has not been considered in the literature, where MLEV-type and XY-type supercycles have been studied separately~\cite{Levitt1982Supercycles, Freeman1982, Gullion1990}.


\section{Discussion and Outlook}
In this work, we introduced U-TRACK, a new family of dynamical decoupling sequences derived from an optimal tracking strategy. The tracking strategy monitors the qubit’s propagator after each pulse block during optimization, enabling dynamic error compensation that suppresses error accumulation block by block. Conventional periodic DD schemes repeat a fixed block, which often leads to systematic error accumulation across repetitions. U-TRACK instead compensates residual errors dynamically at every block, yielding non-periodic pulse schedules that remain robust over long idle times. To efficiently search the space of pulse parameters, we combine GRAPE-based gradient optimization with a genetic algorithm: the genetic algorithm explores the parameter space globally and provides diverse initial guesses, while GRAPE refines them locally to high-fidelity solutions. Furthermore, the optimized sequences exhibit a structured symmetry that can be exploited to warm-start the optimization: using a symmetry-informed initial guess rather than a random one significantly reduces convergence time and improves solution quality. U-TRACK is a general framework: it supports an arbitrary number of blocks $M$ and any even number of pulses $N$ per block, making it applicable across a wide range of hardware constraints and idle-time durations. In this work, we focus on the case $N = 4$ as a concrete demonstration, but the framework extends directly to other configurations.\\

We demonstrated the effectiveness of this framework through the U-TRACK4 sequence, validating its robustness both numerically and experimentally. Experimental validation was performed on a transmon processor at the Walther-Meißner-Institut (WMI), while additional experiments on IQM devices are presented below, demonstrating the cross-platform applicability of U-TRACK sequences. These WMI experiments show substantial improvements in resilience to both static detuning and pulse-amplitude errors. U-TRACK4 maintains high process fidelity over long idle times, whereas XY4 suffers from cumulative residual errors. Simulations further show that UR40 remains sensitive to mid-sequence parameter changes (Sec.~\ref{Sec.Rob_time}). Under random telegraph noise (Sec.~\ref{Sec.Rob_time}), the three sequences perform similarly in the high-frequency regime, where rapid fluctuations average out regardless of sequence structure. In the low-frequency regime and for small detuning amplitudes, however, U-TRACK4 and XY4 maintain significantly higher fidelities than UR40, whose performance degrades severely when mid-sequence parameter changes occur. Within this low-frequency regime, U-TRACK4 further outperforms XY4: when the effective detuning is relatively high, XY4 exhibits sharp fidelity degradation that U-TRACK4 avoids. In addition to these controlled robustness experiments, we performed a dedicated comparison of U‑TRACK4 and XY4 on the IQM Garnet device~\cite{iqm20} under native operating conditions, as detailed in Appendix~\ref{App:IQM_experiment}. These measurements confirm that U‑TRACK4 maintains a clear advantage over XY4 in regimes dominated by pulse distortions and overlap, while providing comparable protection once modest interpulse delays are introduced to mitigate such effects (see Fig.~\ref{fig:DD4_IQM}).\\

It is important to acknowledge that established sequences such as XY64 and MLEV-type supercycles~\cite{Levitt1982Supercycles, Freeman1982, Gullion1990}, which combine cyclic and sign-reversed variants of short base blocks, can achieve performance comparable to U-TRACK4 when applied to sequences of four $\pi$-pulses. This connection is not coincidental: as the analysis in Sec.~\ref{sec:structure} and Appendix~\ref{app:XY64} demonstrates, the optimal tracking algorithm naturally converges to a combination of MLEV-type and XY-type blocks — a structure we term XUR. Notably, to our knowledge, this mixed structure has not been considered in the prior literature, where MLEV-type and XY-type supercycles have been studied separately~\cite{Levitt1982Supercycles, Freeman1982, Gullion1990}. The XUR structure also has practical implications for optimization: using an XUR sequence as a warm-start initial guess leads to faster convergence and improved solution quality compared to random initialization, as discussed in Sec.~\ref{sec:structure}.\\

The U-TRACK framework nevertheless offers several key advantages: (i) it is not limited to $\pi$-gates: the current implementation optimizes phases of $\pi$-rotations, but general pulse-phase shaping could provide additional degrees of freedom to improve robustness and reduce the refocusing period; (ii) it allows flexible control over the number of pulses per block (e.g., U-TRACK4, U-TRACK6, U-TRACK8), whereas XY64 and MLEV sequences are both limited to blocks of 4 pulses and fixed to specific phase patterns; (iii) the optimization can incorporate arbitrary noise models, enabling sequences tailored to device-specific noise spectra; and (iv) the robustness profile can be designed to target specific spectral regions or parameter ranges, as highlighted in previous studies~\cite{kobzar2005pattern, janich2011robust}.\\

Looking forward, several promising directions emerge for extending the U-TRACK framework. The tracking strategy could be generalized beyond single-qubit settings to multi-qubit architectures, where correlated noise, crosstalk, and gate-interaction idling introduce new challenges~\cite{Tripathi2022, Hickman_2026}. A natural starting point is the combinatorial and group-based decoupling constructions developed for interacting spin systems~\cite{santos_advantages_2008}, which could serve as building blocks for tracking-based optimization over multi-qubit blocks. Extending U-TRACK to protect entangled states or idle qubits within larger subroutines may further enhance the stability of noisy quantum processors. In addition, incorporating more advanced pulse-shaping capabilities may enable simultaneous suppression of multiple noise channels, including faster time-varying or spectrally complex noise. Dynamically tailored pulse envelopes, durations, and frequencies—beyond the $\pi$-pulse framework used here—could further expand the practical operating regime of the sequences.\\

The optimization toolkit is released as an open-source Julia package, UTrack.jl~\cite{malvetti2024utrack}, facilitating integration into quantum-control pipelines and experimental workflows. As quantum hardware scales, such dynamically optimized and hardware-agnostic DD sequences may become an essential component of high-fidelity quantum computation.\\

\section{Code Availability}
The optimization toolkit developed in this work is released as an open-source Julia package, UTrack.jl~\cite{malvetti2024utrack}, which supports both CPU and GPU execution for flexible deployment across different computational platforms. The repository includes ready-to-use, pre-optimized U-TRACK sequences with varying pulse-block lengths.
\label{sec:data}

\section{Acknowledgements}
The project is part of the Munich Quantum Valley, which is supported by the Bavarian state government with funds from the Hightech Agenda Bayern Plus.
The authors acknowledge financial support from the German Federal Ministry of Education and Research (BMBF) through the programs "German Quantum Computer based on Superconducting Qubits" (GeQCoS; No. 13N15680) and MUNIQCSC (No. 13N16188), and from the Deutsche Forschungsgemeinschaft (DFG) through the excellence initiative "Munich Center for Quantum Science and Technology" (MCQST; No. 390814868). We thank Niklas Bruckmoser and Julius Feigl for fabricating the superconducting qubit processors and Ivan Tsitsilin for the QPU design used at WMI. AA and IR acknowledge the support of the UK government Department for Science, Innovation and Technology (DSIT) through the UK National Quantum Technologies Programme and the International Science Partnership Fund (ISPF)
\label{sec:acknowledgement}

\bibliography{bib}
\newpage
\appendix
\section{Optimal tracking sequences}
\label{app:optimization}
\subsection{Optimization of U-TRACK sequences} 
\subsubsection{Overview}
The sequence is optimized using a composite pulse made of a concatenation of several $\pi$-rotations with numerically optimized phases $\phi_n$ without inter-pulse delays. One $\pi$-rotation of phase $\phi_n$ is assumed to be of the form:
\begin{equation}
    P^{\epsilon\delta}_n=e^{-i\phi_n\tfrac{\sigma_z}{2}}e^{-i\left(\Omega(1+\epsilon)\tfrac{\sigma_x}{2}+\delta\tfrac{\sigma_z}{2}\right)T_{\pi}}e^{i\phi_n\tfrac{\sigma_z}{2}},
    \label{eq:app_Pn}
\end{equation}
where  $\sigma_{x,y,z}$ are the Pauli matrices, $\Omega$ is the Rabi rate in rad/s, $T_{\pi}=\pi/\Omega$ is the duration of a $\pi$-pulse, $\phi_n$ is the phase of the $n$-th pulse, and where  $\delta$ and $\epsilon$ represent the effects of detuning and pulse amplitude deviations, respectively. After the first DD block, the evolution operator becomes:
\begin{equation}
    U_1^{\epsilon\delta} = P_{N}^{\epsilon\delta}P_{N-1}^{\epsilon\delta}\cdots P_1^{\epsilon\delta},
\end{equation}
where $N$ is the number of $\pi$-pulses in one block. After the second block, the evolution operator is:
\begin{equation}
    U_2^{\epsilon\delta} = P_{2N}^{\epsilon\delta}P_{2N-1}^{\epsilon\delta}\cdots P_{N+1}^{\epsilon\delta} U_1^{\epsilon\delta},
\end{equation}
After $M$ blocks, the operator is:
\begin{equation}
    U_M^{\epsilon\delta} = P_{M\cdot N}^{\epsilon\delta}P_{M\cdot N-1}^{\epsilon\delta}\cdots P_{M\cdot N-N+1}^{\epsilon\delta} U_{M-1}^{\epsilon\delta},
\end{equation}
which is made of $M\times N$ $\pi$-pulses of phases $\phi_{mn}$.

In a standard DD sequence, such as XY4, the pulse phases $\phi_1, \ldots, \phi_N$ are optimized so that, after a single block, the resulting evolution operator satisfies $U^{(1)} \simeq \mathbb{I}$ over a given region of parameter space $(\epsilon,\delta)$. Specifically, this can be achieved by optimizing $\phi_1,\ldots,\phi_N$ to maximize the fidelity:
\begin{equation}
F_1=\eta\int_{\delta_{\min}}^{\delta_{\max}}\int_{\epsilon_{\min}}^{\epsilon_{\max}}\frac{1}{4}\Big|\operatorname{Tr}\left(\mathbb{I}\cdot U_1(\delta,\epsilon)\right)\Big|^2\; d\epsilon\,d\delta,\label{eq:F1}
\end{equation}
where $\eta=1/[(\epsilon_{\max}-\epsilon_{\min})(\delta_{\max}-\delta_{\min})]$ normalizes the fidelity to the area of the parameter region.

The optimized block is subsequently repeated multiple times to span the qubit’s idle period. Consequently, after $M$ blocks, the evolution operator is given by $U_M = U_1^M$. This implies that the fidelities $F_2$, $F_3$, and so forth are fully determined by the first block, and that any residual errors present in this block accumulate over successive repetitions.

In our approach, all blocks are optimized simultaneously by finding a set of pulse phases $\phi_1,\cdots,\phi_{M\times N}$ that maximizes the total fidelity
\begin{equation}
    F=\frac{1}{M}\sum_{m=1}^M F_m\label{eq:TotalFidelity}
\end{equation}
with:
\begin{equation}
F_m=\eta_w\int_{\delta_{\min}}^{\delta_{\max}}\int_{\epsilon_{\min}}^{\epsilon_{\max}}w(\epsilon,\delta)\,\frac{1}{4}\Big|\operatorname{Tr}\left(\mathbb{I}\cdot U_m(\delta,\epsilon)\right)\Big|^2\; d\epsilon\,d\delta,\label{eq:Fm}
\end{equation}
where the weight $w(\epsilon,\delta)$ and its normalization $\eta_w$ are given below.

With this total fidelity being maximized, if residual errors remain after the first block, the subsequent blocks adapt their pulse phases to compensate for them, and this correction process continues throughout the sequence.  This approach maintains a high fidelity across the parameter space even for long idle durations. 

The U-TRACK4 sequence presented in the main text was optimized over $M = 100$ blocks of $N = 4$ 
$\pi$-pulses, with parameter ranges $\delta/(2\pi) \in [-1.5625, 1.5625]$~MHz 
and $\epsilon \in [-0.4, 0.4]$. The maximum detuning corresponds to 
$0.4$ times the Rabi frequency of a $\pi$-pulse of duration 
$T_{\pi} = 128$~ns. Additional weight was assigned to the central region of the parameter space
($\epsilon\sim 0$, $\delta\sim 0$) through a Gaussian factor of the form:
\[w(\epsilon,\delta)=1+w_0\exp\left\{-\frac{1}{2\sigma^2}\left[\left(\tfrac{\epsilon}{\epsilon_{\max}}\right)^2+\left(\tfrac{\delta}{\delta_{\max}}\right)^2\right]\right\}\]
with $w_0=100$ and $\sigma=0.3$. Accordingly, the normalization in Eq.~\eqref{eq:Fm} is $\eta_w=1/\int\!\!\int w(\epsilon,\delta)\,d\epsilon\,d\delta$, which ensures $F_m\leq 1$.

A careful reader may note that we could optimize the fidelity of the form
\begin{equation}
    F_m^{\pm} = \pm \eta_w\int_{\delta_{\min}}^{\delta_{\max}} \int_{\epsilon_{\min}}^{\epsilon_{\max}} w(\epsilon,\delta)\,\frac{1}{2}\,\mathbb{R}\!\left[\operatorname{Tr}\!\left(\mathbb{I}\cdot U_m(\delta,\epsilon)\right)\right]\, d\epsilon\, d\delta, \label{eq:Fmpm}
\end{equation}
where the sign is chosen by the user, instead of Eq.~\eqref{eq:Fm}. The difference is crucial from an optimal-control perspective, as explained below.
In Eq.~\eqref{eq:Fm}, the sign of the identity is not imposed; that is, both $U=-\mathbb{I}$ and $U=+\mathbb{I}$ yield perfect fidelity. By contrast, in Eq.~\eqref{eq:Fmpm} the sign is fixed; for instance, for $F^{+}$ the fidelity is poor when $U=-\mathbb{I}$.
Nevertheless, previous work has shown that the type fidelity given in Eq.~\eqref{eq:Fmpm} converges much more reliably toward the global optimum~\cite{Kobzar2012}, although it is not clear under what circumstances it is preferable to impose the $+$ or $-$ sign.
In our case, since there are many fidelity subfunctions $F_m$, Eq.~\eqref{eq:Fmpm} is not appropriate, as the best choice of sign for each $F_m$ is unclear. It would require exploring two possible choices for each $F_m$, resulting in $2^M$ possible combinations of signs to identify the global optimum.

To optimize the total fidelity in Eq.~\eqref{eq:TotalFidelity}, we use a modified version of the GRAPE algorithm~\cite{Khaneja2005}, a gradient-based method that converges to local optimal solutions near the initial pulse guess. Technical details of the algorithm are provided in Appendix~\ref{subsec:grape_details} below, and an open source package is available online~\cite{malvetti2024utrack}. We observed that the solutions produced by GRAPE strongly depend on the initial pulse guess. This sensitivity becomes more pronounced when optimizing over a larger number of blocks $M$ , suggesting a control landscape with many competing local minima. Finding the best solutions requires optimizing with numerous initial pulse guesses, which can be cumbersome since a single optimization run typically takes a few dozen minutes on a standard laptop. To improve the efficiency of the optimization, we enhanced GRAPE with a global search based on a genetic algorithm (see Sec.~\ref{app:genetic}), which substantially increases the likelihood of obtaining the global optimum, or at least high-quality locally-optimal solutions.

\subsection{Local optimization via gradient descent\label{subsec:grape_details}}
\paragraph{Propagation} 
As explained earlier, the optimization is performed assuming an ensemble of two-level systems of different amplitude errors $\epsilon$ and detuning $\delta$. For a given pair $(\epsilon,\delta)$, the evolution operator after $M$ DD blocks is given by:
\begin{equation}
    \underbrace{P^{\epsilon\delta}_{MN}P^{\epsilon\delta}_{MN-1}\cdots P^{\epsilon\delta}_{2N+1}\underbrace{P^{\epsilon\delta}_{2N}\cdots P^{\epsilon\delta}_{N+1}\underbrace{P^{\epsilon\delta}_N\cdots P^{\epsilon\delta}_1}_{U^{\epsilon\delta}_1}}_{U^{\epsilon\delta}_2}}_{U^{\epsilon\delta}_M},
\end{equation}

\paragraph{Optimized fidelity} In practice, the error parameters $\epsilon$ and $\delta$ are discretized in a grid of $N_\epsilon\times N_\delta$ pairs of values $(\epsilon_k,\delta_{\ell})$, distributed such that $\epsilon_k\in[-\epsilon_{\max},\epsilon_{\max}]$ and $\delta_{\ell}\in[-\delta_{\max},\delta_{\max}]$.
The fidelity of the sequence after $m$ blocks---given by Eq.~\eqref{eq:Fm}---is evaluated using:
\begin{equation}
\begin{aligned}
    F_m & = \frac{1}{W}\sum_{k=1}^{N_\epsilon}\sum_{\ell=1}^{N_\delta} w_{k\ell}\,\frac{1}{4}\Big|\operatorname{Tr}\left(\mathbb{I}\cdot U^{\epsilon_k\delta_{\ell}}_m\right)\Big|^2\\
    &\equiv \frac{1}{W}\sum_{k=1}^{N_\epsilon}\sum_{\ell=1}^{N_\delta} w_{k\ell}\, f^{\epsilon_k\delta_{\ell}}_m,
    \end{aligned}
\end{equation}
representing the weighted average fidelity on the grid, with $w_{k\ell}=w(\epsilon_k,\delta_\ell)$ and $W=\sum_{k,\ell}w_{k\ell}$. The total fidelity that is optimized using optimal tracking is given by:
\begin{equation}
    F=\frac{1}{M}\sum_{m=1}^{M}F_m,
\end{equation}
which corresponds to the average fidelity over all blocks. The goal is thus to find a set of pulse phases $\phi_n$ that maximizes $F$. To do so, we use a gradient algorithm in which gradients are calculated using GRAPE, as explained below.

\paragraph{Gradients}
The gradients correspond to the derivatives of the fidelity with respect to all $M\times N$ phases. They are given by:
\begin{equation}
        \frac{\partial F}{\partial \phi_n} = \frac{1}{M}\sum_{m=1}^{M} \frac{\partial F_m}{\partial \phi_n}.
\end{equation}
Note that since $F_m$ is the fidelity after $m$ blocks, we obtain $\partial F_m/{\partial \phi_n}=0$ if $n>m\times N$ ($N$ is the number of pulses in one block). This means that the above formula can be rewritten as:
\begin{equation}
        \frac{\partial F}{\partial \phi_n} = \frac{1}{M}\sum_{m\geq \tfrac{n}{N}}^{M} \frac{\partial F_m}{\partial \phi_n}.
\end{equation}
Continuing to apply the chain rule, we get:
\begin{equation}
    \begin{aligned}
        \frac{\partial F}{\partial \phi_n}=\frac{1}{MW}\sum_{m\geq \tfrac{n}{N}}^{M}\sum_{k=1}^{N_\epsilon}\sum_{\ell=1}^{N_\delta} w_{k\ell}\,\frac{\partial f^{\epsilon_k\delta_{\ell}}_m}{\partial \phi_n}.
    \end{aligned}
\end{equation}
Since $f^{\epsilon_k\delta_{\ell}}_m$ is determined by the evolution operators $U^{\epsilon_k\delta_{\ell}}_m$, we can use the chain rule~\cite{MatrixCookBook}:
\begin{equation}
    \frac{\partial f^{\epsilon_k\delta_{\ell}}_m}{\partial \phi_n} = \mathbb{R}\left[\Tr\left(\frac{\partial f^{\epsilon_k\delta_{\ell}}_m}{\partial U^{\epsilon_k\delta_{\ell}}_m}\frac{\partial U^{\epsilon_k\delta_{\ell}}_m}{\partial\phi_n}\right)\right].
    \label{eq:app_dfdUdUdp_0}
\end{equation}
The first term is a derivative of a scalar with respect to a matrix, leading to:
\begin{equation}
    \frac{\partial f^{\epsilon_k\delta_{\ell}}_m}{\partial U^{\epsilon_k\delta_{\ell}}_m}=\frac{1}{2}\Tr\big(U^{\epsilon_k\delta_{\ell}}_m\big)^{*}\,\mathbb{I}.
    \label{eq:app_dfdU}
\end{equation}
The second term is given by:
\begin{equation}
    \frac{\partial U^{\epsilon_k\delta_{\ell}}_m}{\partial\phi_n} = P^{\epsilon_k\delta_{\ell}}_{mN}\cdots P^{\epsilon_k\delta_{\ell}}_{n+1}\frac{\partial  P^{\epsilon_k\delta_{\ell}}_{n}}{\partial\phi_n}P^{\epsilon_k\delta_{\ell}}_{n-1}\cdots P^{\epsilon_k\delta_{\ell}}_1.
    \label{eq:app_dUdphin}
\end{equation}
The term in the middle is derived using Eq.~\eqref{eq:app_Pn}, which leads to:
\begin{equation}
    \frac{\partial  P^{\epsilon_k\delta_{\ell}}_{n}}{\partial\phi_n} = -i\left[\frac{\sigma_z}{2},P^{\epsilon_k\delta_{\ell}}_{n}\right],
    \label{eq:appdPndphin}
\end{equation}
where $[\cdot,\cdot]$ represents the Lie product. Using equations~\eqref{eq:appdPndphin},~\eqref{eq:app_dUdphin} and~\eqref{eq:app_dfdU}, we obtain:
\begin{equation}
\resizebox{\linewidth}{!}{$
    \begin{aligned}
    &\frac{\partial f^{\epsilon_k\delta_{\ell}}_m}{\partial \phi_n} = \\
    &\frac{1}{2}\mathbb{R}\Bigg[\Tr\big(U^{\epsilon_k\delta_{\ell}}_m\big)^{*}\,\Tr\Bigg(\underbrace{P^{\epsilon_k\delta_{\ell}}_{mN}\cdots P^{\epsilon_k\delta_{\ell}}_{n+1}}_{\Lambda^{\epsilon_k\delta_{\ell}}_{m,n+1}}\left[\frac{-i\sigma_z}{2},P^{\epsilon_k\delta_{\ell}}_{n}\right]\underbrace{P^{\epsilon_k\delta_{\ell}}_{n-1}\cdots P^{\epsilon_k\delta_{\ell}}_1}_{\hat{U}^{\epsilon_k\delta_{\ell}}_{n-1}}\Bigg)\Bigg].
    \label{eq:app_dfdUdUdp_1}
    \end{aligned}$
    }
\end{equation}
Here, $\hat{U}^{\epsilon_k\delta_{\ell}}_{n-1}$ represents the evolution operator obtained after applying the first $n-1$ $\pi$-pulses. The operator $\Lambda^{\epsilon_k\delta_{\ell}}_{m,n+1}$ is often referred to as \emph{costate} or \emph{adjoint state}, and is obtained using a backward propagation. 

\subsection{Global optimization via genetic algorithm}
\label{app:genetic}
While the GRAPE algorithm explained in the previous section can efficiently compute the gradient of the total fidelity $F$ with respect to the pulse phases $\phi_n$, and hence allows for implementing a gradient-based local optimization, it turns out that the cost function contains many local minima that are far from being global minima. Hence, the final result strongly depends on the initial pulse. A possible approach is to use many random initial pulses, optimize them all using GRAPE, and simply choose the best final pulse. Using the problem structure stemming from tracking fidelity, we can improve on this naive approach.

The idea is to use a genetic algorithm, in conjunction with GRAPE, to optimize a \emph{population} of pulses over several \emph{generations}. Note that while genetic algorithms have been used previously in quantum control and dynamical decoupling~\cite{devra2018efficient,Lidar_GADD}, combining them with local optimization via GRAPE significantly improves the results. The general structure of the algorithm is as follows:

\paragraph{Genetic algorithm}

\begin{enumerate}[(1)]
\item Generate a population of $J$ random initial pulses $\tilde\phi_n^{(0,j)}$ with $j=1,\ldots,J$. 
\item Locally optimize each pulse using GRAPE to obtain the pulses $\phi_n^{(0,j)}$. This is the $0$-th generation.
\item For $i=1,\ldots,I$:
    \begin{enumerate}
    \item Apply a genetic update to the population $\phi_n^{(i-1,j)}$ to obtain $\tilde\phi_n^{(i,j)}$.
    \item Locally optimize all pulses using GRAPE to obtain $\phi_n^{(i,j)}$.
    \end{enumerate}
\end{enumerate}

Now, let us explain the genetic update. This step is based on the symmetry properties explained in Sec.~\ref{sec:structure}.

\begin{enumerate}
\item \emph{Elitism:} Keep the best performing pulses. This guarantees that the best fidelity of each generation does not get worse.
\item \emph{Mutation:} Some pulses are randomly modified in the following ways:
    \begin{enumerate}
    \item Some of the blocks are permuted randomly.
    \item Some blocks are modified by flipping, reversing, or cyclically permuting the phases (note that we don't include a phase shift since this is a continuous parameter which is taken care of by the GRAPE algorithm)
    \end{enumerate}
\item \emph{Recombination:}
We randomly choose pairs of pulses and swap some of their blocks.
\end{enumerate}

The key property of the genetic update is that a good pulse remains good under mutation, and two good pulses remain good under recombination. In this way, the algorithm does not search the entire space of all possible pulses, as happens with random pulses, but rather focuses on a subset of pulses that all perform well. As a result, the cost function improves significantly over the course of several generations, and the final optimized pulses obtain a nice structure as illustrated in Figure~\ref{fig:pulses}.

\subsection{Highly parallel implementation}
The optimization algorithm described in the previous two sections lends itself to parallelization. The GRAPE's gradient computation is performed in two steps. A forward and a backward propagation. The same operation must be performed for each value of the amplitude error and detuning $(\epsilon_k,\delta_\ell)$. Using a GPU, all of these can be performed in parallel, and the final contributions to the gradient are simply added up. Moreover, GRAPE can be applied to many pulses within a population in parallel, as the computations are independent. 
\section{Experimental Robustness Characterization of DD Sequences
\label{App:WMI_experiment}}
This appendix describes the experimental implementation of the dynamical decoupling (DD) sequences studied in this work, as well as the robustness characterization measurements performed on a transmon qubit processor fabricated at the Walther-Meißner-Institut (WMI). The fidelity of each sequence was assessed via quantum process tomography (QPT)~\cite{NielsenChuangBook}.

The DD sequences were implemented on a superconducting transmon qubit device~\cite{Koch-Transmon}. In particular, we used one qubit of a chip containing 17 qubits and 24 couplers in a square-grid architecture, with qubit frequency $\omega_q/2\pi = \SI{5497.798 \pm .002}{\mega\hertz}$. The coherence parameters were $T_1 = \SI{19.9 \pm 2.8}{\micro\second}$, $T_2^{\text{Echo}} = \SI{37 \pm 8}{\micro\second}$, and $T_2^{\text{R}} = \SI{29 \pm 6}{\micro\second}$. The experimental setup and wiring follow the description in~\cite{Glaser-closed-loop}.

The qubit drive was generated using an SHFSG from Zurich Instruments~\cite{ZurichInstruments_SHFSG_UserManual}.
Each $\pi$ pulse was implemented using a $4\sigma$ Gaussian envelope with a duration of $\SI{128}{\nano\second}$, modulated at the driven qubit frequency. The $\pi$-pulse amplitude and drive frequency were calibrated using error amplification sequences.

The sequences were implemented via calibrated $\pi$-$X$ rotations, with appropriate pre- and post-oscillator phase increments.
This realizes the desired equatorial-axis $\pi$ rotation via $X_\theta \equiv R_z(\theta)\,X\,R_z(-\theta)$, where $X \equiv R_x(\pi)$.
Deviations from an ideal $X$ rotation must be accounted for in the DD sequence. Consequently, the simplified rearrangement $R_z(\theta)\, X\, R_z(-\theta) \propto X\, R_z(-2\theta)$, which is valid only for a perfect $X$ gate up to a global phase, is in general not applicable in the presence of pulse imperfections.

On the FPGA-based control instrument, only a single waveform is stored, and the oscillator phase increments are applied by iterating consecutive command table entries.

To reconstruct the experimental process implemented on the device, we performed single-qubit quantum process tomography.
We prepared the four input states $|0\rangle$, $|1\rangle$, $|+\rangle = (|0\rangle + |1\rangle)/\sqrt{2}$, and $|-i\rangle = (|0\rangle - i|1\rangle)/\sqrt{2}$.
For each input state, we measured expectation values in the $x$, $y$, and $z$ bases by applying post-rotation gates.
Using these measurements, we reconstructed the quantum channel matrix following the standard protocol described in Ref.~\cite{NielsenChuangBook} and computed the process fidelity~\cite{johnston2011quantum}.
No additional post-processing was performed after QPT.
Therefore, the differences in fidelity color intensity between experiment and simulation in Fig.~\ref{fig:FidelityMaps} directly reflect the underlying process fidelities and are not influenced by any post-processing corrections.

To measure sequence detuning, the I-Q envelope of the Gaussian pulses was modulated at the desired offset frequencies, and corresponding oscillator phase shifts were applied.
Static amplitude shifts were realized by scaling the pulse envelope.
The pre- and post-rotations used for process tomography were performed using the calibrated gates.

\section{Simulation of transmon qubits\label{app:simulations}}
\subsection{Evolution operator}
In the methods described in Appendix~\ref{app:optimization}, Eq.~\eqref{eq:app_Pn} represents the evolution of a two-level system driven by a constant $\pi$-pulse subject to detuning and amplitude errors. In practice, however, all DD sequences in this work are implemented on transmon-based quantum processors, where the $\pi$-pulses have a finite duration and a smooth envelope, and where higher-energy levels of the transmon can be populated. Consequently, while Eq.~\eqref{eq:app_Pn} provides a reasonable approximation of the dynamics, it does not fully capture effects arising from the pulse shape or from the presence of higher excited states.

To accurately reproduce the experimental results, we simulate the qubit dynamics under the pulse shape used in the experiment by numerically solving the time-dependent Schrödinger equation, 
\begin{equation}
    \dot{\tilde{P}}(t) = -i H(t) \tilde{P}(t), \label{eq:SchrodTransmon}
\end{equation}
where $\tilde{P}$ is the evolution operator and $H(t)$ is the Hamiltonian in the rotating frame under the rotating-wave approximation:  
\begin{equation}
    H(t) = \tfrac{1+\epsilon}{2}\big( \Omega(t)a^{\dagger} + \Omega^{*}(t)a \big)
      + \delta\,a^{\dagger}a
      + \tfrac{\alpha}{2}\,a^{\dagger 2}a^{2}.
      \label{eq:TransmonHamiltonian}
\end{equation}
Here, $\Omega(t) = \Omega_I(t) + i\,\Omega_Q(t)$ is the complex control pulse, with $\Omega_I(t)$ and $\Omega_Q(t)$ denoting the in-phase and quadrature components, respectively. The parameter $\alpha$ is the transmon anharmonicity, with $\alpha/(2\pi) = -173~\mathrm{MHz}$, and $a$ ($a^{\dagger}$) are the annihilation (creation) operators. The pulse envelope is the one used in the experiment (see Appendix~\ref{App:WMI_experiment}), namely a truncated Gaussian pulse without quadrature component:  
\begin{equation}
    \begin{cases}
        \Omega_I(t) = A\left[e^{-(t-T_{\pi}/2)^2/(2\sigma^2)} - e^{-T_{\pi}^2/(8\sigma^2)}\right],\\[6pt]
        \Omega_Q(t) = 0,
    \end{cases}
    \label{eq:Simu_GaussianPulse}
\end{equation}
for $0 \le t \le T_{\pi}$, where $T_{\pi} = 128~\mathrm{ns}$ is the pulse duration and $\sigma = T_{\pi}/4$. The constant offset ensures that the envelope vanishes at $t=0$ and $t=T_{\pi}$. The amplitude $A$ is set by the condition $\int_0^{T_{\pi}}\Omega_I(t)\,\mathrm{d}t = \pi$, such that the pulse implements a $\pi$ rotation when $\delta = \epsilon = 0$. This gives
\begin{equation}
    A/(2\pi)=8.44~\mathrm{MHz}.
    \label{eq:GaussianAmp}
\end{equation}
In our simulations, the pulse is assumed to be piecewise constant with time steps of $0.5~\mathrm{ns}$, and we truncate the operators $a$ and $a^{\dagger}$ to the first three energy levels. This yields a $3\times 3$ propagator, $\tilde{P}(\delta,\epsilon,T_{\pi})$, representing the numerical solution of Eq.~\eqref{eq:SchrodTransmon} under the pulse of Eq.~\eqref{eq:Simu_GaussianPulse}. The propagator corresponding to a $\pi$-rotation with phase $\phi_n$ is then given by
\begin{equation}
    \tilde{P}_n(\delta,\epsilon) = e^{-i\phi_n a^{\dagger}a}\,\tilde{P}(\delta,\epsilon,T_{\pi})\,e^{i\phi_n a^{\dagger}a}.\label{eq:PropagatorTransmon}
\end{equation}
Thus, after $M$ cycles of $N$ pulses, the evolution operator of the system is a $3\times 3$ matrix given by:
\begin{equation}
    \tilde{U}_M(\delta,\epsilon) = \tilde{P}_{M\cdot N}(\delta,\epsilon)\tilde{P}_{M\cdot N-1}(\delta,\epsilon)\cdots \tilde{P}_1(\delta,\epsilon).
\end{equation}
With this operator, the fidelity at $(\delta,\epsilon)$ is computed in the computational subspace as:
\begin{equation}
    f(\delta,\epsilon) = \frac{1}{4}\Big|\bra{0}\tilde{U}_M(\delta,\epsilon)\ket{0}+\bra{1}\tilde{U}_M(\delta,\epsilon)\ket{1}\Big|^2.\label{eq:FidTransmon}
\end{equation}
Population leaked to $\ket{2}$ reduces this fidelity.
As shown in Fig.~\ref{fig:FidelityMaps}, this technique provides an excellent match with the experimental data.

\section{Comparison with MLEV64 and XY64\label{app:XY64}}
In this section, we compare U-TRACK4 to two existing sequences from the literature built from short blocks applied with phase shifts 
between blocks: MLEV64~\cite{Levitt1982Supercycles} and 
XY64~\cite{Gullion1990,lancaster2025}. All sequences are truncated to 40 
$\pi$-pulses (10 blocks of 4 $\pi$-pulses) to match the same idle 
time used throughout the paper. Simulations are performed on a transmon-type Hamiltonian, where 
each $\pi$-pulse is implemented as a Gaussian pulse, as described in Appendix~\ref{app:simulations}.\\

\begin{figure*}[t!]
\includegraphics[width=1\textwidth]{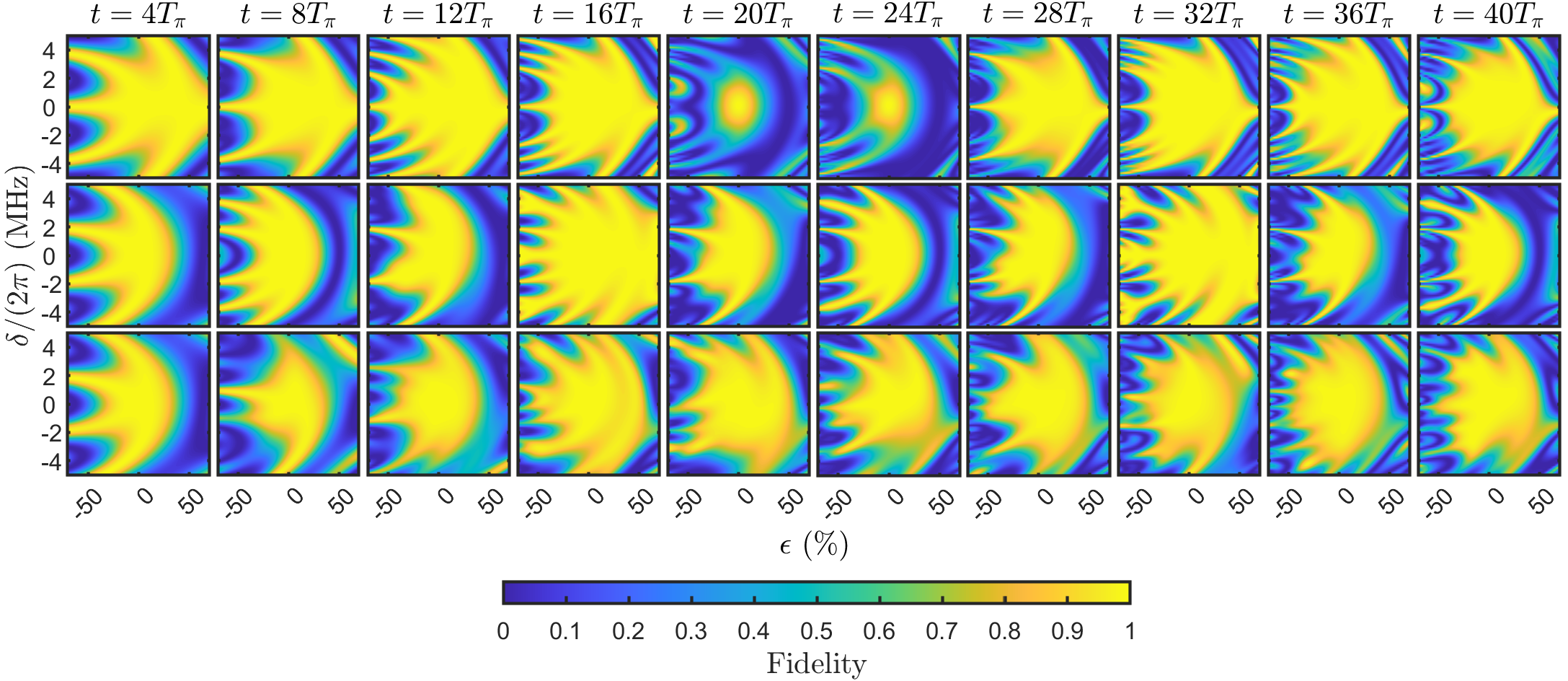}
\caption{Robustness profile of MLEV64 (top row), XY64 (middle row), and U-TRACK4 (bottom row), as functions of pulse-amplitude error $\epsilon$ and frequency detuning $\delta$. Each column corresponds to the fidelity 
after a given number of blocks of 4~$\pi$-pulses, as indicated at the 
top. Each $\pi$-pulse has a duration of $T_{\pi} = 128$~ns. All sequences are truncated to 40 
$\pi$-pulses (10 blocks of 4 $\pi$-pulses). \label{fig:MLEV64}}
\end{figure*}

\begin{figure}
    \includegraphics[width=0.65\linewidth]{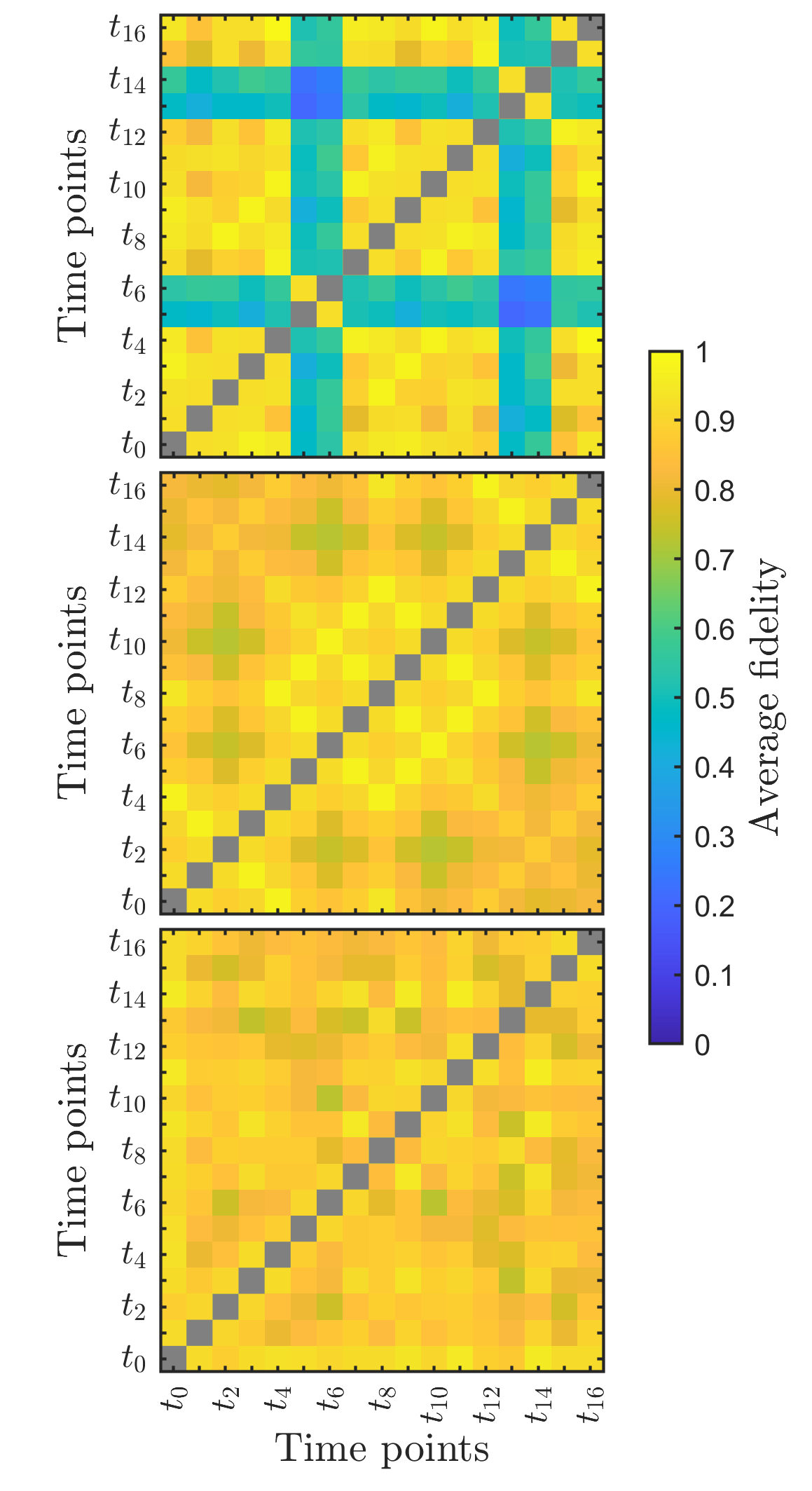}
    \caption{Average fidelity of the identity gate as a function of pulse-segment boundaries for MLEV64 (top), XY64 (middle), and U-TRACK4 (bottom). All sequences here are shown for 64 
$\pi$-pulses (16 blocks of 4 $\pi$-pulses). The time coordinate $t_n$ denotes the time after $n$ blocks of four $\pi$-pulses. Each pixel at coordinates $(t_m, t_n)$ displays the average fidelity $\bar{F}_{mn}$ (Eq.~\ref{eq:avgfidpixel}), which quantifies how well the pulse segment $(t_m, t_n)$ implements the identity gate despite frequency-offset and amplitude errors. Trivial diagonal entries corresponding to $U({t_{n},t_{n}}) = \mathbb{I}$ are grayed out.\label{fig:figureMatrixMLev}}
\end{figure}

As for the robustness profile across a range of parameter deviations shown in Fig.~\ref{fig:FidelityMaps}, we plot a simulated robustness profile for MLEV64, XY64, and U-TRACK4 sequences in Fig.~\ref{fig:MLEV64}, showing the fidelity after each block of 4 
$\pi$-pulses. Overall, all three sequences achieve comparable fidelity profiles. However, MLEV64 shows a noticeable degradation at $t = 20T_{\pi}$ and $t = 24T_{\pi}$, which directly impacts the results of 
Fig.~\ref{fig:figureMatrixMLev}, as discussed below. XY64 does not suffer from this issue and maintains performance comparable to U-TRACK4 across all 10 blocks.\\     

To provide a better overview, we reproduce Fig.~\ref{fig:FigIntro}, this time extending to 16 blocks (i.e., 64 pulses), for the three sequences mentioned above. The results are shown in Fig.~\ref{fig:figureMatrixMLev}, where each pixel of coordinate (n,m) shows the average fidelity over a range of robustness: 
\begin{equation}
\begin{split}
    \bar{F}_{nm}=\eta\int_{\delta_{\min}}^{\delta_{\max}}\int_{\epsilon_{\min}}^{\epsilon_{\max}}&\frac{1}{4}\Big|\bra{0}U_{\epsilon\delta}(t_n, t_m)\ket{0}\\&+\bra{1}U_{\epsilon\delta}(t_n, t_m)\ket{1}\Big|^2\; d\epsilon\,d\delta,
\end{split}
\label{eq:avgfidpixel}
\end{equation}
where, as in Eq.~\eqref{eq:FidTransmon}, the fidelity is evaluated in the computational subspace, $\eta=1/[(\epsilon_{\max}-\epsilon_{\min})(\delta_{\max}-\delta_{\min})]$, and $U_{\epsilon\delta}(t_n, t_m)$ represents the evolution operator between time points $t_n$ and $t_m$. In other words, it measures how well the identity gate performs on the pulse segment between $t_n$ and $t_m$. We take the average over 
$[\delta_{\min}/(2\pi),\delta_{\max}/(2\pi)] \in [-1.5625, 1.5625]$~MHz (optimized range)
and $\epsilon \in [-0.4, 0.4]$.

The MLEV64 profile exhibits prominent blue lines, indicating a sharp drop in fidelity at specific time points --- notably at $t_5$ ($t = 20\, T_{\pi}$), $t_6$ ($t = 24\, T_{\pi}$), $t_{13}$, and $t_{14}$. These correspond to pulse segments that fail to implement the identity gate robustly; this can also be observed directly in the robustness plots at $t = 20\, T_{\pi}$ and $t = 24\, T_{\pi}$ of Fig.~\ref{fig:MLEV64}. Consequently, the average fidelity over the chosen parameter range is significantly degraded at these points.
\section{Experimental comparison of DD sequences}
\label{App:IQM_experiment}
To experimentally evaluate the performance of the U‑TRACK4 and XY4 dynamical decoupling (DD) sequences, we conducted a targeted experiment designed to quantify their ability to preserve a specific quantum state. In analogy to the procedure used for the results presented in Fig.~\ref{fig:FidelityMaps}, we performed full quantum process tomography following the methodology detailed in Appendix~\ref{App:WMI_experiment}. In contrast to the experiments in Sec.~\ref{Sec.Robust_para}, where amplitude deviations ($\epsilon$) and static detuning ($\delta$) were intentionally introduced to test robustness, here no such modifications were applied, and the sequences were tested under the system’s natural conditions. 

\begin{figure}
 \centering
 \includegraphics[width=0.25\textwidth]{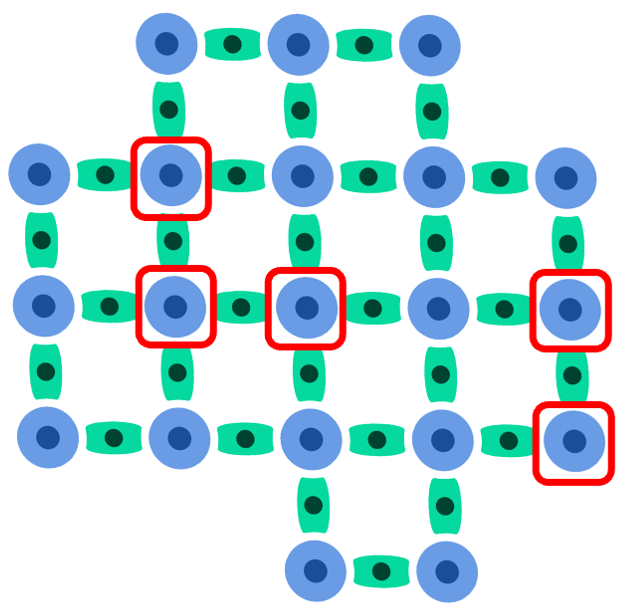}
 \caption{Topology of the 20‑qubit IQM Garnet device. The five qubits enclosed by the red box indicate the subset used in the experiments.}
 \label{fig:iqm_qpu}
\end{figure}

\begin{figure}
 \centering
 \includegraphics[width=0.45\textwidth]{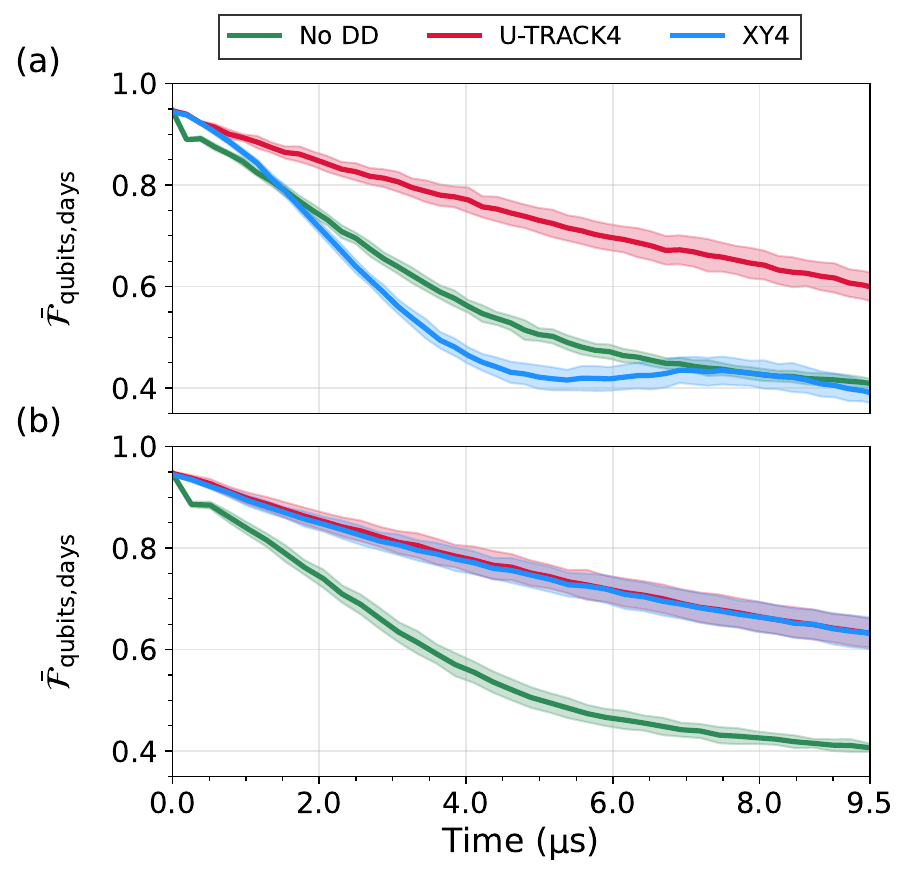}
 \caption{Performance comparison of U‑TRACK4 and XY4 dynamical decoupling sequences. (a) Process fidelity $\mathcal{F}$ as a function of total sequence duration without inter‑pulse delays. (b) Process fidelity $\mathcal{F}$ as a function of total sequence duration with inter‑pulse delays inserted between $\pi$ pulses. In both panels, solid lines show the average fidelity $\mathcal{\bar{F}}$ over qubits and measurement days, and shaded regions indicate the corresponding standard deviation across days.}
 \label{fig:DD4_IQM}
\end{figure}

All experiments were performed on the 20‑qubit superconducting device IQM Garnet~\cite{iqm20}, in which both computational qubits and tunable couplers are implemented as flux‑tunable transmon qubits. For this study, we selected five qubits on the device, as indicated in Fig.~\ref{fig:iqm_qpu}. Each single‑qubit gate has a duration of $24~\mathrm{ns}$. Full quantum process tomography was performed simultaneously on all five qubits. Specifically, we prepared the four input states ${|0\rangle, |1\rangle, |+\rangle, |\!-\!i\rangle}$, applied the DD sequences with varying numbers of cycles, and measured expectation values in the $x$, $y$, and $z$ bases. In the ``No‑DD'' case, no gates were applied, and the qubits evolved freely for the same duration as the corresponding DD pulse block. The experiment was repeated across five different days (spanning approximately eight calibration cycles). Fig.~\ref{fig:DD4_IQM} reports the mean and standard deviation of the process fidelity $\mathcal{F}$ as a function of time, averaged over all qubits and days.

The top panel of Fig.~\ref{fig:DD4_IQM} shows the results obtained when the DD pulses were applied without any inter‑pulse delays. In this regime, the XY4 sequence performed worse than applying no DD at all. This behavior is likely caused by pulse distortions that lead to partial overlap between successive pulses when the pulse intervals are too short, resulting in shifts in the effective rotation axis~\cite{Sheldon2016, Hyyppa2024}. In contrast, the U‑TRACK4 sequence demonstrated robustness to such distortions and exhibited significantly improved performance. We also observed similar effects on IBM quantum devices. A detailed investigation of these effects lies beyond the scope of this work.

We repeated the experiment with an added inter‑pulse delay of $\tau = 40~\mathrm{ns}$ between each DD pulse ($P$), with the resulting sequence taking the form $P_{1}-\tau-P_{2}-\tau-P_{3}-\tau-P_{4}-\tau$. The corresponding results are shown in the bottom panel of Fig.~\ref{fig:DD4_IQM}. In this case, the performance of U‑TRACK4 and XY4 was comparable, indicating that both sequences provide similar protection of quantum states when pulse overlap effects are mitigated. These results constitute preliminary observations; a more detailed experimental analysis will be presented in future work.

\end{document}